\documentclass[12pt,twoside,chap,a4paper]{article}
\usepackage{jheppub}
\pdfoutput=1
\usepackage{graphicx}
\usepackage{amssymb,latexsym} %Fonts
\usepackage{graphics,subfigure,graphicx}
\usepackage{amsmath,xspace}
     
\usepackage{eepic}
\newcommand{\Herwig}{\textsf{Herwig}\xspace}

\usepackage{array}
\newcolumntype{L}[1]{>{\raggedright\let\newline\\\arraybackslash\hspace{0pt}}m{#1}}
\newcolumntype{C}[1]{>{\centering\let\newline\\\arraybackslash\hspace{0pt}}m{#1}}
\newcolumntype{R}[1]{>{\raggedleft\let\newline\\\arraybackslash\hspace{0pt}}m{#1}}

\title{
Logarithmic Accuracy of Angular-Ordered Parton Showers
}
\author[a]{Gavin~Bewick}
\author[a]{Silvia~Ferrario~Ravasio}
\author[a,b]{Peter~Richardson}
\author[c]{Michael~H.~Seymour}
\affiliation[a]{Institute for Particle Physics Phenomenology, Durham University}
\affiliation[b]{Theoretical Physics Department, CERN, Switzerland}
\affiliation[c]{Lancaster-Manchester-Sheffield Consortium for Fundamental Physics, School of Physics and Astronomy, University of Manchester, M13 9PL, U.K.}
\emailAdd{gavin.bewick@durham.ac.uk}
\emailAdd{silvia.ferrario-ravasio@durham.ac.uk}
\emailAdd{peter.richardson@durham.ac.uk}
\emailAdd{Michael.Seymour@manchester.ac.uk}
\abstract{We study the logarithmic accuracy of angular-ordered parton showers by considering the singular limits
          of multiple emission matrix elements. This allows us to consider different choices for the
          evolution variable and propose a new choice which has both the correct logarithmic behaviour and improved
          performance away from the singular regions. In particular the description
          of $e^+e^-$ event shapes in the non-logarithmic region is significantly improved.}

\preprint{\begin{flushright}MAN/HEP/2019/003\\CERN-TH-2019-047\\MCnet-19-08\\IPPP/19/30\end{flushright}}

\begin{document}

\maketitle
\flushbottom

\section{Introduction}

Monte Carlo event generators~\cite{Bellm:2017bvx,Bellm:2015jjp,Sjostrand:2014zea,Gleisberg:2008ta},
which provide a complete description of the complicated hadronic final state observed
in high-energy particle collisions, are essential tools as their results can be directly compared with experimental
measurements. These simulations combine a calculation of the hard scattering process, usually at next-to-leading order accuracy,
with parton shower~(PS) evolution from the scale of the hard process to a low energy scale where non-perturbative
hadronization models describe the formation of hadrons from the quarks and gluons of the perturbative calculation. Together with
a non-perturbative model of multiple parton scattering and decay of the primary hadrons, these generators simulate the final
hadronic state.\footnote{For a complete review of the approximations and models used see ref.\,\cite{Buckley:2011ms}.}

Most of the progress made in this field over the last decade came from matching the parton shower approximation of QCD radiation with fixed-order matrix elements. This increased the accuracy of the cross-section calculation and improved the description of hard radiation, which is not adequately described by the soft and collinear approximations used in parton shower algorithms.  In the last few years however there has been a revival
of work~\cite{Kato:1986sg,Kato:1988ii,Kato:1990as,Kato:1991fs} to improve the accuracy of the parton shower algorithm
in antenna~\cite{Giele:2011cb,Hartgring:2013jma,Li:2016yez} and dipole~\cite{Hoche:2017hno,Hoche:2017iem,Dulat:2018vuy} showers,
as well as work on amplitude-based evolution to treat subleading colour effects~\cite{Platzer:2012np,Martinez:2018ffw}.

A recent work~\cite{Dasgupta:2018nvj} showed that two popular dipole
shower algorithms, used in \textsf{PYTHIA 8}~\cite{Sjostrand:2004ef}
and \textsf{Dire}~\cite{Hoche:2015sya}, have issues even at
leading-logarithmic accuracy due to the way the singular emissions are
split between different dipole contributions and how recoils are
handled.  The authors considered an initial $q\bar{q}$ dipole and the
emission of two gluons $g_1$ and $g_2$ that are both soft and
collinear to either of the hard partons and widely separated in
rapidity from each other. Given these requirements, the two emissions
must be independent and the double-emission probability is
\begin{equation}
d P^{(2)}_{\rm soft} = \frac{1}{2!}\prod_{i=1}^2 \left[C_{\rm F}\frac{\alpha_s(p_{T i})}{\pi}\,\frac{d\phi_i}{2\pi}\, \frac{d p_{T i}^2}{p_{T i}^2}\, d y_i\right],
\label{eqn:doubleSoftEmissionProb}
\end{equation}
where $y_i$ is the rapidity of the gluon $i$ and $p_{T i}$ is
its transverse momentum, all computed in the original $q\bar{q}$
dipole frame, where the $z$ axis is aligned with the $q$ direction.
The second gluon, $g_2$, can be emitted either from the $\bar{q}-g_1$
or from the $q-g_1$ dipole.  However, although $g_2$ may be
further from $g_1$ than $g_1$ is from $q$ or $\bar{q}$,
when the event is looked at in the emitting-dipole frame, $g_2$ may be 
closer in angle to $g_1$, which will thus
play the role of the emitter. This results in an 
incorrect colour factor, since $C_{\rm A}/2$ is assigned instead of 
$C_{\rm F}$. This mistake has no effect at leading colour, since 
$C_{\rm F} \to C_{\rm A}/2$ in the large number of colours limit, 
though it does 
correspond to an error in the subleading colour contribution. 
Furthermore, if $g_1$ is identified as the emitting particle 
in the emitting dipole, it has to balance the transverse momentum 
of $g_2$ and
\begin{equation}
\mathbf{p}_{T 1} \to \mathbf{p}_{T 1} - \mathbf{p}_{T 2},
\end{equation}
where the bold symbol indicates it is a two-momentum.
This implies that $p_{T 1}$ can receive a substantial modification if the
transverse momentum of the second gluon is only marginally smaller than
that of the first emission, thus violating Eqn.~\eqref{eqn:doubleSoftEmissionProb}. 

In this paper we will use a similar approach to that of
Ref.\,\cite{Dasgupta:2018nvj} in order to analyse the behaviour of the
improved angular-ordered shower of Ref.~\cite{Gieseke:2003rz}.
The subleading colour issue does not affect an angular-ordered parton shower,
which implements colour coherence by construction, so
that in the above example $g_2$ can only be emitted, with the correct colour factor, in a cone
around $q$ or $g_1$ that is smaller than the angle that separates
$q$ and $g_1$.
However, the effect of the recoil must be carefully taken into
account. The angular-ordered parton shower, which uses a ``global''
recoil (the momenta of all partons in the shower are changed to ensure
momentum conservation) and $1\to2$ splittings, is significantly
different from the dipole showers, which implement ``local'' recoil
(where only the momenta of colour-connected partons change to ensure
momentum conservation), as considered in
Ref.~\cite{Dasgupta:2018nvj}. While some of the issues
considered in Ref.~\cite{Dasgupta:2018nvj} are irrelevant for parton
showers using $1\to2$ kinematics and global recoil, some of the
underlying physics issues addressed can occur in the angular-ordered
parton shower, although they manifest themselves in different ways.

In the next section we briefly introduce the relevant features
of a massless parton shower algorithm, including a definition of logarithmic accuracy which will guide our analysis.
In section~\ref{sec:kinematics} we present the definitions of the parton momenta and kinematics used in the angular-ordered parton shower.
These are then used to construct three different interpretations of the evolution variable and consider the logarithmic accuracy of each.
We then discuss the tuning procedure used for the \textsf{Herwig~7} 
angular-ordered parton shower to ensure a like-for-like comparison between
new and old evolution variables. Finally we present our conclusions. In Appendix~\ref{app:gtoqqbar} we discuss a technical detail related to the splitting $g\to q \bar{q}$ and in Appendix~\ref{sec:thrustLogs} we explicitly show that the current default recoil scheme implemented in \textsf{Herwig 7} only correctly describes the double-logarithmically enhanced terms, thus justifying the proposal of a new recoil prescription.

\section{Definition of Logarithmic Accuracy}
\label{sec:log_counting}

Fixed-order calculations quickly become cumbersome when we increase the particle multiplicity to take into account the emission of extra jets.
However, the leading contribution from such emissions arises in the soft and collinear regions of the phase space, \emph{i.e.}\ when we consider the emission of a gluon with vanishing energy or of a parton whose momentum is parallel to the momentum of the emitter.
In this latter limit the cross section for the emission of an extra parton is fully factorised, so that we can easily derive the emission probability
\begin{equation}
d P = \frac{\alpha_s}{2 \pi} \frac{d t}{t} d z P(z),
\label{eq:Pcollinear}
\end{equation}
where $z$ is the light-cone momentum fraction (see eq.~\eqref{eqn:zdef}), $P(z)$ are the collinear splitting functions and $t$ is a scale that approaches 0 in the collinear limit. 
We see that if we try to integrate the collinear emission probability in \eqref{eq:Pcollinear} over the available phase space,
there is a logarithmic divergence for $t \to 0$.
When we consider the emission of a soft gluon, \emph{i.e.}\ when $z \to 1$, we have another source of logarithmic singularities as the  splitting kernels all behave like
\begin{equation}
\lim_{z \to 1} P(z) = \frac{2 C}{1-z},
\end{equation}
where $C=C_A$ in case of gluon splitting and $C=C_F$ if the gluon is emitted from a quark line.
This simple approximation allows us to correctly take into account double logarithms associated with soft-collinear gluon emission and single logarithms associated with a collinear branching.

When we consider $n$ branchings, we can have at most $2 n$ large logarithms, $L$, of widely disparate scales of the problem, which arise if all the emissions are simultaneously soft and collinear: this means that the emission probability is proportional to $ \alpha_s^n L^{2n}$, and we call such contributions leading logarithms~(LL). 
The use of quasi-collinear splitting functions \cite{Catani:2000ef} gives the first next-to-leading~(\emph{i.e.}\ single) collinear logarithms (NLL), \emph{i.e.}\ $ \alpha_s^n L^{2n-1}$, and together with the choice of 
the two loop running coupling evaluated using the Catani-Marchesini-Webber scheme~\cite{Catani:1990rr} at the transverse momentum of the radiated partons~\cite{Catani:1989ne}, includes all leading~(double) and next-to-leading~(single) logarithmic contributions,  except for those due to soft wide angle gluon emissions.

In general defining a strict logarithmic accuracy for a parton shower algorithm is difficult. Formally a parton shower algorithm has only leading logarithmic accuracy, although it is able to capture many next-to-leading contributions. There are some classes of infrared-safe observables where an improved coherent branching formalism leads to full next-to-leading log accuracy (\emph{e.g.}\ in semi-inclusive hard processes such as deep inelastic scattering and Drell-Yan at large $x$ \cite{Catani:1990rr}). 
In Ref.~\cite{Dasgupta:2018nvj} it was shown that in some regions of the phase space the double-soft-gluon emission probability is not correctly described by dipole showers. In practice, neglecting subleading colour contributions\footnote{
In Ref.~\cite{Baberuxki:2019ifp} resummed predictions at NLO+NLL accuracy are compared against dipole shower predictions for the case of 4-, 5- and 6-jet Durham resolutions to assess the impact of subleading colour contributions. In the (strict) large number of colours (LC) approximation significant differences are found, however the colour treatment of parton showers (that associates $C_A/2$ when a gluon emission come from a gluon leg, $C_F$ from a quark leg) leads to results almost identical to those obtained considering the full-colour dependence.}, the parton shower approximation of eq.~\eqref{eqn:doubleSoftEmissionProb} fails only when the transverse momenta of the two emitted gluons are commensurate and thus the recoil procedure quite significantly changes the transverse momentum of the first emission. Since logarithms of commensurate scales are small, it was also found that, for a wide range of event-shape  observables, the leading terms are correct but the next-to-leading logarithmic terms are wrong.

Based on this observation, a necessary (but not sufficient) condition for an algorithm to be next-to-leading log accurate is that the singularity structure of the spectrum in Eqn.~\eqref{eqn:doubleSoftEmissionProb} is reproduced in all the regions of the Lund plane \cite{Andersson:1988gp}, which describes the available phase space in terms of the transverse momenta and rapidities of the emitted gluons relative to a suitably-defined frame/axis.
As was first pointed out in Ref.~\cite{Andersson:1988gp}, and exploited in Ref.~\cite{Dasgupta:2018nvj} to understand the logarithmic accuracy of parton showers, the leading-logarithmic gluon emission is uniform in the plane defined by the logarithm of this transverse momentum and rapidity. Specific corrections to the uniform distribution can be made in specific phase space regions, to promote this description to next-to-leading logarithmic. In more detail, as the cut-off of a parton shower, or value of an event shape observable, is made logarithmically smaller ($\mathcal{O}< ^{-L}$), the area of the Lund plane increases as the square of this logarithm, $\sim L^2$. If a parton shower algorithm makes an order~1 error over an \emph{area} of the Lund plane, \emph{i.e.}~a region that grows at rate proportional to $L^2$, we say that it is not leading-logarithmically accurate. Conversely, if it does not make such an error, we say that it has the potential to be leading-logarithmically accurate. If a parton shower algorithm makes an order~1 error only along a \emph{line} in the Lund plane, \emph{i.e.}~a region that grows at rate proportional to $L$, we say that it is leading-logarithmically accurate but not next-to-leading-logarithmically accurate. Our aim is to construct an algorithm that makes order~1 errors only at isolated \emph{points} in the Lund plane, \emph{i.e.}~regions that do not grow with $L$, and therefore give rise only to errors in event shape distributions of either next-to-next-to-leading logarithmic or power-suppressed order. Emission of two gluons of similar transverse momenta corresponds to a line in the Lund plane and therefore careful consideration of this configuration is required to reach next-to-leading logarithmic accuracy. The importance of recoil effects for correct description of this region was first pointed out in Ref.~\cite{Andersson:1991he}.

In the following we will consider three recoil scheme prescriptions, one of which leads to an incorrect kinematic mapping in the soft limit. In Appendix.~\ref{sec:thrustLogs} we explicitly show how this leads to incorrect NLL contributions in the thrust distribution as an example event shape observable.

\section{Kinematics}
\label{sec:kinematics}

We will define all momenta in terms of the Sudakov basis such that the 4-momentum
of particle $l$ is
\begin{equation}
  q_l = \alpha_l p + \beta_l n +k_{\perp l},
  \label{eqn:sudakov}
\end{equation}
where the reference vectors $p$ and $n$ are the momentum of the parent parton with on-shell mass $m_0$ and a lightlike vector that points in the direction of its colour partner. They obey 
  \begin{equation}
p^2=m_0^2, \qquad p\cdot n \neq 0, \qquad
n^2=0, \qquad 
p\cdot k_{\perp l} = n\cdot k_{\perp l}=0,
  \end{equation}
so that the transverse momenta are defined with reference to the direction of $p$ and $n$ and the transverse momentum 4-vector $k_{\perp l}$ is spacelike. 
If we consider a particle $\widetilde{ij}$ that splits into a pair of particles $i$ and $j$, the light-cone momentum fractions of particles $i$ and $j$ are defined as
  \begin{equation}
    z_i= \frac{q_i \cdot n }{q_{\widetilde{ij}} \cdot n}=\frac{\alpha_i}{\alpha_{\widetilde{ij}}}=1-z_j.
    \label{eqn:zdef}
  \end{equation}
The relative transverse momentum of the branching is given by
\begin{equation}
q_{\perp i} \equiv k_{\perp i} - z_i k_{\perp \widetilde{ij}} = k_{\perp j} - z_j k_{\perp \widetilde{ij}},
\end{equation}
and the magnitude of the spatial component is therefore given by
\begin{equation}
p^2_{Ti} \equiv \mathbf{p}^2_{\perp i} = -q^2_{\perp i}.
\end{equation}
The parton shower evolution terminates when 
\begin{equation}
p_{T i}^2 < p_{T \min}^2,
\end{equation}
where $p_{T \min}^2$ is an infrared cutoff tuned to data of the order of 1~GeV.

For many results we will not need a specific representation of the reference vectors.
If we do need a representation we will use the choice made in Ref.\,\cite{Gieseke:2003rz} for final-state radiation with a final-state colour partner,
\emph{i.e.}
\begin{subequations}
  \begin{align}
    p &= \frac{Q}2\left[1+b-c,0,0,\lambda \right]; \\
    n &= \frac{Q}2\left[\lambda,0,0,-\lambda \right];
  \end{align}\end{subequations}
where $Q$ is the invariant mass of the radiating particle and its colour partner, \mbox{$b=m^2_0/Q^2$}, \mbox{$c=m^2_s/Q^2$},
$\lambda$ is the K\"{a}ll\'{e}n function 
\begin{equation}
\lambda = \lambda(1,b,c) \equiv \sqrt{1+b^2+c^2-2b-2c-2bc},
\label{eqn:Kallen}
\end{equation}
and $m_0$, $m_s$ are the masses of the radiating particle
and its colour partner, respectively.

\subsection{Single Emission}

For the branching $0\to 12$, with no further emission we have:
\begin{subequations}
\begin{eqnarray}
  q_0&=&  p + \beta_0 n; \\
  q_1&=& z p +\beta_1 n + q_{\perp};    \\
  q_2&=& (1-z)p +\beta_2 n - q_{\perp};
\end{eqnarray}
\end{subequations}
where, $q_\perp$ is the transverse momentum 4-vector, $m_{0,1,2}$ are
the on-shell masses of the particles, $z$ is the light-cone momentum
defined in Eqn.~\ref{eqn:zdef}, $\beta_{1,2}$ are determined by the
on-shell condition $q_{1,2}^2=m_{1,2}^2$ and $\beta_0$ by momentum
conservation.  The virtuality of the parton initiating the branching
is therefore
\begin{equation}
  q_0^2 = \frac{p_T^2}{z(1-z)}+ \frac{m_1^2}{z}+\frac{m_2^2}{(1-z)},\label{eqn:virt1}
\end{equation}
where $q_\perp^2=-p_T^2$.

\subsection{Second emission} 

\begin{figure}
  \begin{center}
    \includegraphics[width=0.5\textwidth]{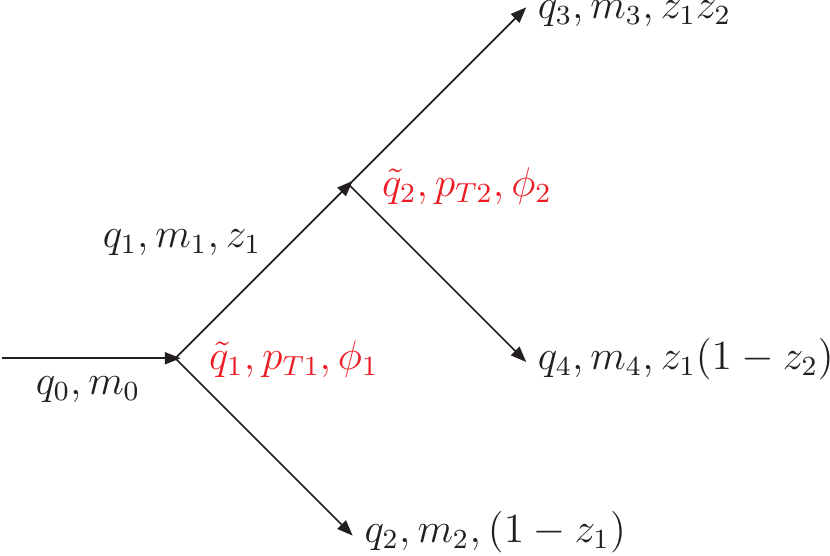}
  \end{center}
  \caption{The kinematics of two branchings in the angular-ordered parton shower. The off-shell momenta~($q_i$),
    on-shell masses~($m_i$) and light-cone momentum fractions of the partons are shown together with
    the evolution variable~($\tilde{q}_i$), transverse momentum~($p_{Ti}$) and azimuthal angle~($\phi_i$) of each branching.}
  \label{fig:kinematics}
\end{figure}

We now consider two emissions, the first with $z_1$, $\tilde{q}_1$, $\phi_1$
and the second from the first outgoing parton of the first branching with
$z_2$, $\tilde{q}_2$, $\phi_2$, as shown in Fig.\,\ref{fig:kinematics}.

We define the off-shell momenta of the four partons after the branchings as:
\begin{subequations}
  \begin{eqnarray}
    q_0 &=& p +\beta_0 n;\\
    q_1 &=&    z_1 p +\beta_1 n + q_{\perp 1}; \\
    q_2 &=& (1-z_1)p +\beta_2 n - q_{\perp 1}; \\
    q_3 &=& z_1z_2 p +\beta_3 n + z_2q_{\perp 1} +q_{\perp 2};\\
    q_4 &=& z_1(1-z_2)p+\beta_4 n + (1-z_2)q_{\perp1}-q_{\perp2};
\end{eqnarray}\end{subequations}
where $p^2=m^2_0$,
the $\beta_i$ coefficients are fixed by the on-shell condition and momentum
conservation and the space-like transverse momentum
\begin{equation}
  q_{\perp i} = \left[ 0; {\bf p}_{Ti},0\right] = \left[0; p_{Ti}\cos\phi_i,p_{Ti}\sin\phi,0\right], 
\end{equation}
such that $q_{\perp i}^2=-{\bf p}_{Ti}^2=-p_{T i}^2$.
The virtualities of the branching partons are:
\begin{subequations}
  \begin{eqnarray}
    q^2_0 &=& \frac{p_{T1}^2}{z_1(1-z_1)}  +\frac{q^2_1}{z_1}+\frac{m_2^2}{1-z_1}; \\
    q^2_1 &=& \frac{p_{T2}^2}{z_2(1-z_2)}  +\frac{m^2_3}{z_2}+\frac{m^2_4}{1-z_2}.
  \end{eqnarray}\label{eqn:virtualities}\end{subequations}

In all the cases we will consider parton $4$ will be a gluon, $m_4=0$, so that partons $1$ and $3$ must 
have the same mass, \emph{i.e.}\ $m_1=m_3$.
It will also prove useful to define a unit vector in the direction of the transverse momentum, \emph{i.e.}
\begin{equation}
  {\bf \hat{n}}_i = \left[\cos\phi_i,\sin\phi_i \right].
  \label{eqn:nhat}
\end{equation}

\section{Interpretation of the Evolution Variable}

In Ref.~\cite{Gieseke:2003rz} the extension of the original angular-ordered
parton shower~\cite{Marchesini:1983bm} to include mass effects and longitudinal boost invariance along the
jet axis was presented.
In this algorithm the evolution variable is
\begin{equation}
  \tilde{q}^2 = \frac{q_0^2-m_0^2}{z(1-z)} ,
  \label{eqn:qtilde_defn0}
\end{equation}
in order to include mass effects, in particular the correct mass in the propagator,
retain angular-ordering and have a simple single emission probability
\begin{equation}
{\rm d} \mathcal{P} = \frac{{\rm d}\tilde{q}^2}{\tilde{q}^2}\frac{\alpha_S}{2\pi}
\frac{{\rm d} \phi}{2\pi} {\rm d} z P_{i\to jk}(z,\tilde{q}),
\label{eqn:prob}
\end{equation}
where $P_{i\to jk}(z,\tilde{q})$ is the quasi-collinear splitting function \cite{Catani:2000ef},
$z$ is the light-cone momentum fraction
and $\phi$ is the azimuthal angle of the transverse momentum
generated in the splitting. The strong coupling $\alpha_S$ is evaluated at the scale
\begin{equation}
\mu = z(1-z) \tilde{q};
\label{eqn:alphaScale}
\end{equation}
from Eqns.~\eqref{eqn:qtilde_defn0} and~\eqref{eqn:virt1} we can see
that $\mu$ coincides with the transverse momentum of the
splitting~\cite{Catani:1990rr,Bahr:2008pv}, which we label $p_T$, if $m_1=m_2=0$.

For a single emission (or the last emission in an extended shower) where
the children are on their mass-shell, the kinematics are unambiguously
defined by Eqn.\,\ref{eqn:qtilde_defn0} and the ordering variable can be expressed 
equivalently in terms of $q^2$ and $p_T^2$:
 \begin{equation}
  \tilde{q}^2 = \frac{q_0^2-m_0^2}{z(1-z)}= \frac{p_T^2+(1-z)m_1^2+zm_2^2-z(1-z)m_0^2}{z^2(1-z)^2}.
  \label{eqn:qtilde_defn0_long}
\end{equation}
However, when the children of a branching go on to branch further so
that they are off-shell, it is clear from Eqn.~\eqref{eqn:virtualities}
that we cannot preserve simultaneously $q_0^2$ and $p_T^2$.  The
choice of the preserved quantity will determine the interpretation
of~$\tilde{q}^2$.  The procedure used by \Herwig{} is to first
generate a value of $\tilde{q}^2$, $z$ and $\phi$ for a branching and
calculate the preserved kinematic variable from them. 
Then the upper limit
of $\tilde{q}^2$ is calculated for each of the children and the shower proceeds to the next branching. Only at the end of the whole shower evolution, is the
generation of each branching completed by constructing its
kinematics from its (now off-shell) children's momenta, using the
kinematic variable that had been constructed from~$\tilde{q}^2$. Thus
any other kinematic variables are shifted slightly, to accommodate the
change from on-shell to off-shell kinematics.  The interested reader
can find further details concerning the kinematic reconstruction in
Sec.~6.1 of Ref.~\cite{Bahr:2008pv}.  As the virtuality acquired from
the new partons does not depend upon the azimuthal angle, as can be
seen from Eqn.~\eqref{eqn:virtualities}, we can already anticipate that
the shift in the other kinematic variables is not affected by the
value of $\phi$.

We will investigate three different choices for the kinematic variable
that is \mbox{preserved}.

\subsection[$p_T$ preserving scheme]{\boldmath{$p_T$} preserving scheme}

The original choice of Ref.\,\cite{Gieseke:2003rz} was to use Eqn.\,\ref{eqn:qtilde_defn0} together with the expression of the virtuality in Eqn.\,\ref{eqn:virt1}, to define the transverse momentum of the branching $0\to 12$,

\begin{equation}
    p_T^2 = z^2(1-z)^2\tilde{q}^2+m_0^2z(1-z)-m_1^2(1-z)-m_2^2z, \label{eqn:qtilde_defn1}
\end{equation}
where on-shell masses $m_{1,2}$\footnote{By default a cut-off on the transverse momentum of the splitting is applied, as described at the beginning of Sec.~\ref{sec:kinematics}. However it is possible to choose a cut-off on the virtuality of the emitting parton: if this choice is adopted, $m_{1,2}$ are set to the value of the minimum virtualities allowed for particles $1$ and $2$.} are used for the particles produced in the branching.

As observed in Ref.\,\cite{Reichelt:2017hts} this choice tends to give too much hard radiation in the parton shower, as the virtuality of the parent parton can arbitrarily grow after multiple emissions.

\subsection[$q^2$ preserving scheme]{\boldmath{$q^2$} preserving scheme}
  
  Ref.\,\cite{Reichelt:2017hts} suggested that the virtuality of the
  branching should be determined using the virtualities the particles produced in the branching develop after subsequent evolution, such that
  \begin{equation}
    p_T^2 = z^2(1-z)^2\tilde{q}^2+m_0^2z(1-z)-q_1^2(1-z)-q_2^2z. \label{eqn:qtilde_defn2}
  \end{equation}
  Clearly this is the same as Eqn.\,\ref{eqn:qtilde_defn1} if there is no further emission, \emph{i.e.}\
  \mbox{$q^2_{1,2}=m^2_{1,2}$}.

  This choice, however, has the problem that the subsequent evolution of the partons is not guaranteed to result in
  a physical, \emph{i.e.}\ a $p_T^2\geq0$, solution of Eqn.\,\ref{eqn:qtilde_defn2}. In Ref.\,\cite{Reichelt:2017hts}
  it was noted that the vetoing of emissions that give a non-physical solution affected the logarithmic evolution of
  the total number of particles, \emph{i.e.}\ the leading-logarithmic evolution was not correct. Hence, if there was no physical solution
  the transverse momentum was set to zero such that the virtuality of the branching particle is
  \begin{equation}
    q_0^2 = \frac{q_1^2}{z}+\frac{q_2^2}{(1-z)}.
  \end{equation}

  We remark that, even if the transverse momentum $p_T$ of the previous
  emission changes, the strong coupling of that splitting remains
  evaluated at $z_1 (1-z_1) \tilde{q}_1$, \emph{i.e.}\ the original transverse
  momentum in case of massless splitting. Analogously, each emission
  can be vetoed only when it is generated, so subsequent emissions
  will not affect this veto.

\subsection{Dot-product preserving scheme}
  
      Motivated by the original massless
      angular-ordered parton shower of Ref.\,\cite{Marchesini:1983bm}, where the
      evolution variable was related to the dot product of the outgoing momenta, we investigate the choice
  \begin{equation}
    \tilde{q}^2 = \frac{2q_1\cdot q_2 +m_1^2+m_2^2-m_0^2}{z(1-z)},
  \end{equation}
  where the inclusion of the masses is required to give the correct propagator in the general case.
  However, it is not needed for gluon emission, $m_0=m_1$ and $m_2=0$, and only becomes relevant in $g\to q \bar{q}$ branching.
  
  In this case
  \begin{equation}
    p_T^2 = z^2(1-z)^2\tilde{q}^2 -q_1^2(1-z)^2 -q_2^2z^2 +z(1-z)\left[m^2_0-m_1^2-m_2^2\right].\label{eqn:qtilde_defn3}
  \end{equation}
  As before this reduces to the same result in the case of no further emission.

  The major advantage of the original massless algorithm~\cite{Marchesini:1983bm} was
  that the subsequent evolution would always have a physical solution for the transverse momentum.
  If we consider gluon emission the condition
\begin{equation}
  \tilde{q}^2 > 2\max\left(\frac{q_1^2}{z^2},\frac{q_2^2}{(1-z)^2}\right),
  \label{eqn:qtilde3_ineq}
\end{equation}
 is sufficient, but not necessary, for there to be a solution for the transverse momentum in Eqn.\,\ref{eqn:qtilde_defn3}.

 If this inequality is satisfied, the virtuality of the branching parton is
\begin{equation}
 q_0^2 = q_1^2 + q_2^2 + z(1 - z)\tilde{q}^2 \leq \frac{\tilde{q}^2}2.
\end{equation}
Assuming that the branching parton was produced in a previous branching,
with light-cone momentum fraction $z_i$ and evolution scale $\tilde{q}_i$,
the angular-ordering condition ensures that $\tilde{q} < z_i \tilde{q}_i$.
Hence
 \begin{equation}
q_0^2 \leq \frac{z_i^2\tilde{q}^2_i}2,
 \end{equation}
 so that if Eqn.\,\ref{eqn:qtilde3_ineq} is satisfied for one branching
 it will also be satisfied for previous branchings.
 So provided that we require 
\begin{equation}
  \tilde{q}^2 > 2\max\left(\frac{m_1^2}{z^2},\frac{m_2^2}{(1-z)^2}\right),
  \label{eqn:qtilde3_ineq2}
\end{equation}
where $m_{1,2}$ are either the physical, or cut-off masses of the partons, the subsequent evolution
will be guaranteed to have  a physical solution for the transverse momentum.

There are two issues with this choice. The first is that if we impose Eqn.\,\ref{eqn:qtilde3_ineq2}
on radiation from a heavy quark with mass $m$, the transverse momentum of the branching must satisfy
\begin{equation}
  p_T\geq (1-z)m,
\end{equation}
which, since $p_T\sim(1-z)E\theta$ corresponds to $\theta\ge m/E$,
\emph{i.e.}\ the hard dead-cone~\cite{Marchesini:1989yk,Dokshitzer:1991fd} the new algorithm was designed to avoid~\cite{Gieseke:2003rz}.
In practice we use a cut-off on the transverse momentum of the emission which is fine for radiation from gluons and
light quarks, and also for the charm quark since the cut-off is close to the charm mass.
For the 3\textsuperscript{rd} generation quarks we get a small fraction of events where the kinematics cannot be
reconstructed~($\lesssim0.2$ per mille and $\lesssim0.5\%$ of $q\to qg$ branchings for bottom and top quarks, respectively, hardly varying with centre-of-mass energy).
However this region is subleading, \emph{i.e.}\ does not give rise to either soft or collinear logarithms,
and therefore we adopt the approach of setting the transverse momentum of the emission to zero as above in this case.

The second, although less important, issue is the $g\to q \bar{q}$ branching. The limit in this case is presented in Appendix~\ref{app:gtoqqbar}. 
For massive quarks, in particular the bottom quark, this limit is stricter than the cut-off on the transverse momentum we use.
We therefore have some $g\to b\bar{b}$ branchings where we are forced to set the transverse momentum to zero. Again this region is
subleading~($\lesssim0.5\%$ of $g\to b\bar{b}$ branchings, again hardly varying with centre-of-mass energy) and therefore does not affect the logarithmic accuracy. In this case the  $g\to q \bar{q}$ only gives logarithms of the quark mass, and the neglected region does not contribute to these
logarithms.

A full study of these mass effects is beyond the scope of this work, although
very important and we hope to return to it in the future.

\subsubsection{Phase-space corrections}
\label{sec:veto}

The angular ordering of the parton shower, which allows a consistent treatment
of colour coherence effects, leads to regions of phase space without any
gluon emissions. This is the so-called \emph{dead zone}.

\begin{figure}[tb]
\centering
\includegraphics[width=0.45\textwidth]{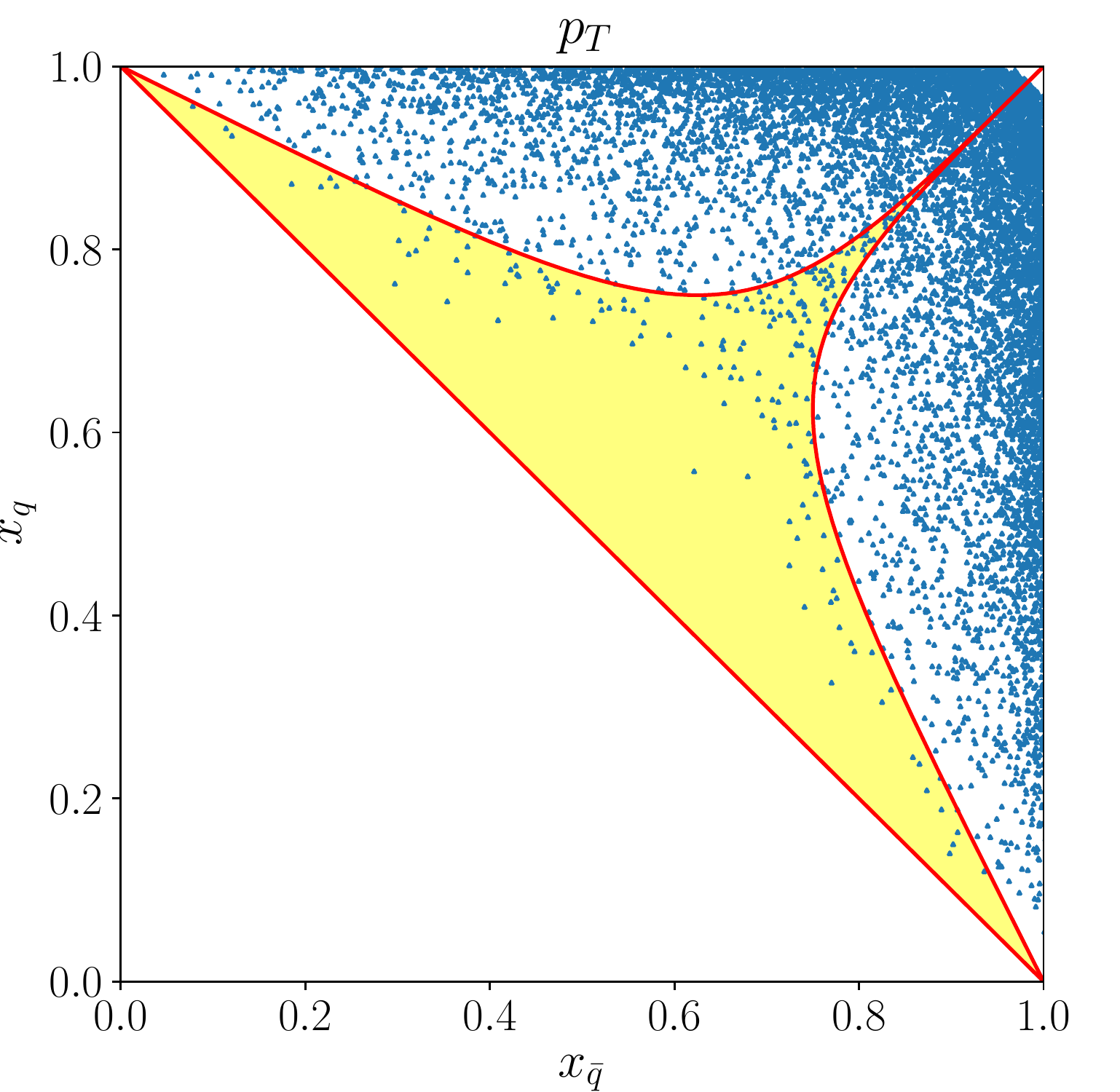}
\includegraphics[width=0.45\textwidth]{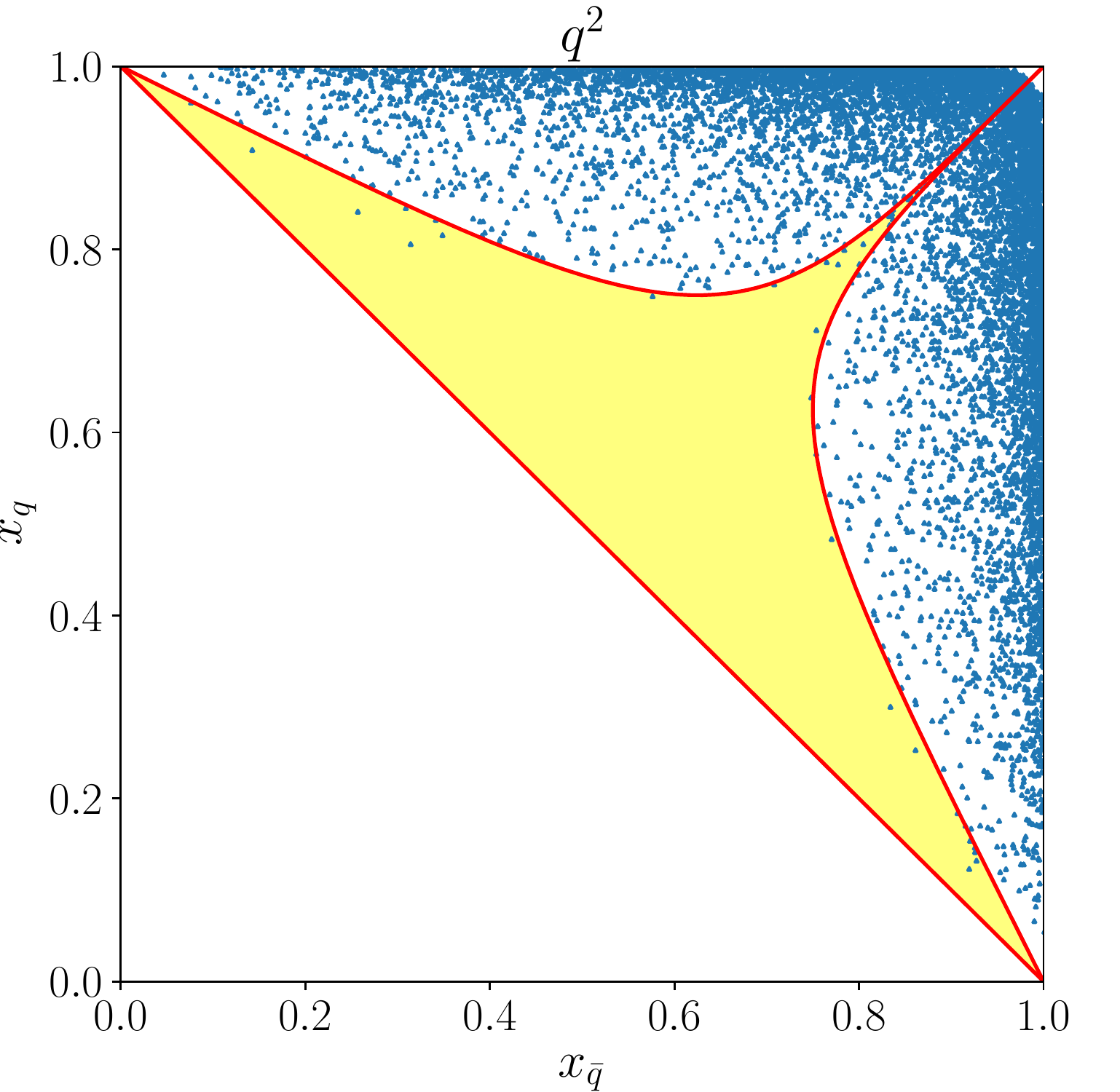}\\
\includegraphics[width=0.45\textwidth]{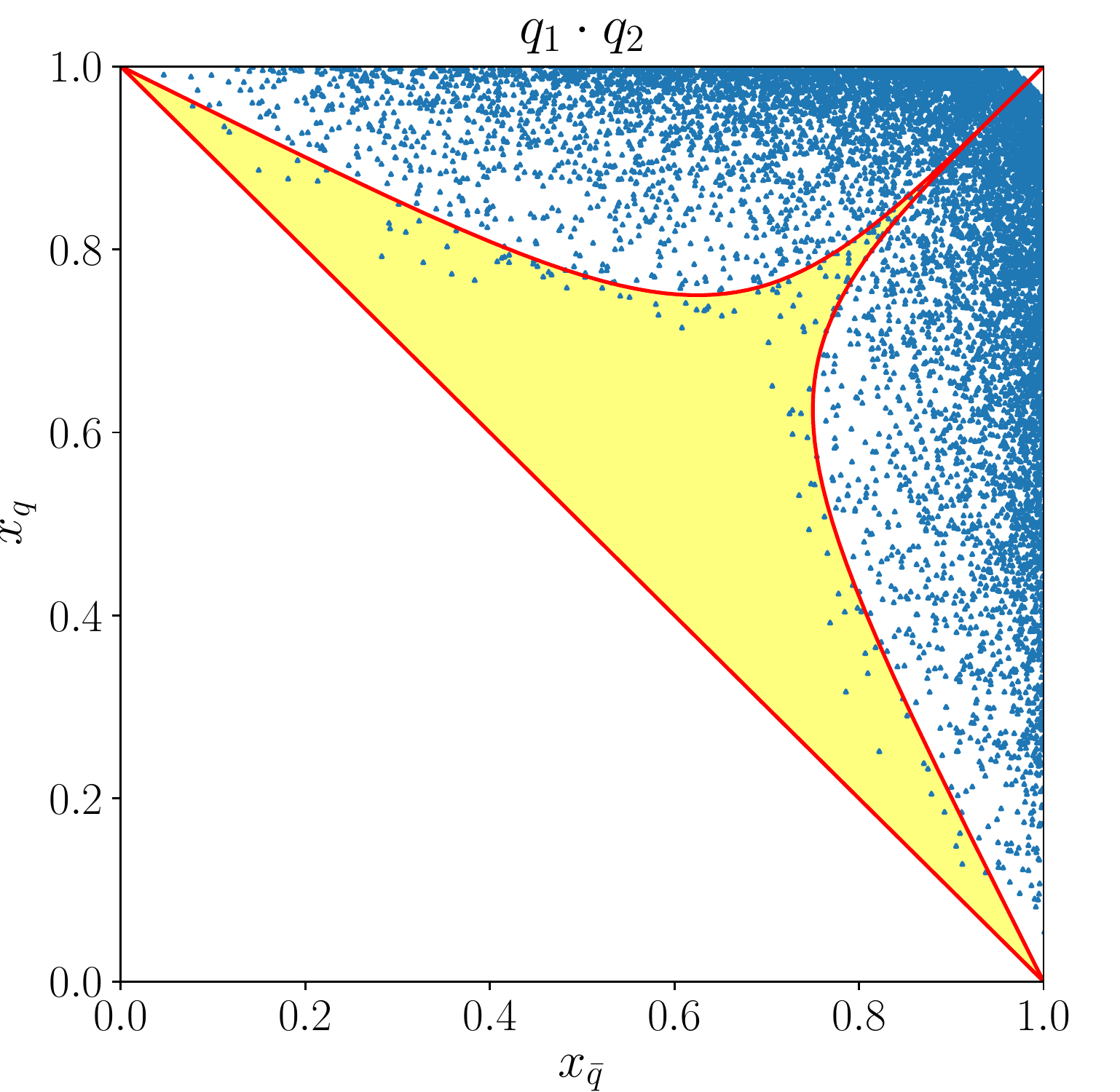}
\includegraphics[width=0.45\textwidth]{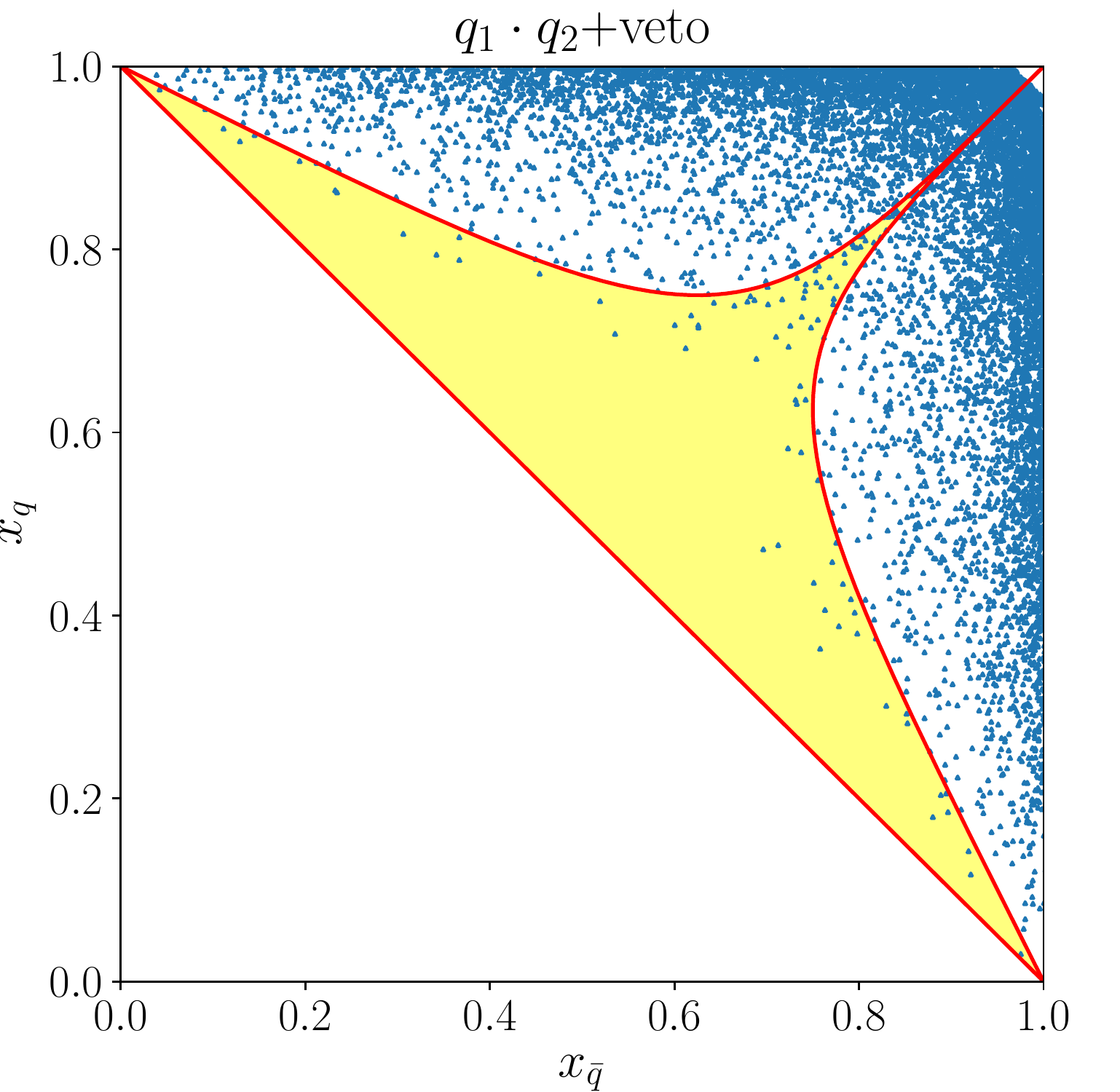}
\caption{Dalitz  plot  for $e^+e^- \to q\bar{q}$ showing  the  region  of  phase
space filled after multiple emission from the quark and anti-quark in
the angular-ordered parton shower for several choices of the preserved
quantity: $p_T$~(upper-left pane), $q^2$~(upper-right pane),
dot-product~(lower-left pane) and  dot-product plus $q^2$ veto~(lower-right
pane). 
The red line illustrates the limits for the first parton-shower emission and the yellow region corresponds to the \emph{dead zone}. The variable $x_i$ is defined to be $2E_i/Q$, where $E_i$ is the energy of parton $i$ and $Q$ is the total energy, all defined in the centre-of-mass of the collision. }
\label{fig:dalitz}
\end{figure}

The choice of the preserved quantity in the presence of
multiple emissions can significantly affect the phase-space region that is
filled by the shower.  Fig.~\ref{fig:dalitz} illustrates the Dalitz plot
for $e^+e^- \to q \bar{q}$. We have clustered the partons using the
FastJet~\cite{Cacciari:2011ma} implementation of the $k_T$ jet
algorithm~\cite{Catani:1991hj} and we have switched off $g\to q\bar{q}$
splittings in order to unambiguously define the $q$ and $\bar{q}$ jets.  We can appreciate how little the
$q^2$-preserving scheme populates the \emph{dead zone}, coloured in yellow, in opposition to the
$p_T$-preserving scheme. This feature is essential when matching to
higher order computations, like matrix element corrections,
since they will take care to fill this hard region of the phase space.
We notice that the dot-product-preserving scheme (bottom-right pane) displays an
intermediate behaviour between the two older schemes, with the number of points in the
dead zone for the dot-product-preserving scheme about half of that in $p_T$-preserving scheme.

In order to enforce the similarities between the dot-product preserving
scheme and the $q^2$ one, that is the current \Herwig{} default, we
implemented a rejection veto to avoid generating too large virtualities.
Indeed the virtuality of the shower progenitor, \emph{i.e.}\ the emitter
particle that was present prior to the shower, increases when multiple
emissions are generated, only in the $q^2$-preserving scheme is it kept
fixed. To this end, let us consider the two-body phase space for the process
$e^+e^- \to q\bar{q}$, which reads
\begin{equation}
d \Phi_2(s, m^2, m^2) = \frac{d \Omega}{32\pi^2} \lambda\!\left(1,
\frac{m^2}{s}, \frac{m^2}{s} \right),
\end{equation}
where $\Omega$ is the solid angle that describes the direction of the quark and 
$\lambda$ is the K\"{a}ll\'{e}n function introduced in Eqn.~\eqref{eqn:Kallen}.
When $n$ emissions are generated the phase space becomes
\begin{equation}
d \Phi_{n+2} = d \Phi_2(s, k^2_q, k^2_{\bar{q}}) \prod_{i=1}^n \frac{d
  \tilde{q}^2_i}{(4\pi)^2}\,z_i(1-z_i)d z_i\,\frac{d\phi_i}{2\pi},
\end{equation}
where $k^2_l$ is the virtuality developed by the shower progenitor $l=q,\bar{q}$.
Thus, if we want to factorize the phase space over the original two-body one, we
need to include the Jacobian factor
\begin{equation}
J = \frac{d \Phi_2(s, k^2_q, k^2_{\bar{q}}) }{d \Phi_2(s, m^2, m^2)} = \frac{\lambda(s,k^2_q,k^2_{\bar{q}})}{\lambda(s,m^2,m^2)}.
\end{equation}
Since $J<1$, we can simply implement a reweighting procedure: at the end of
the showering phase we generate a 
random number $r$ smaller than 1 and we accept the event only if $r<J$,
otherwise we shower the event anew.
Looking at the Dalitz plots (bottom panel of Fig.~\ref{fig:dalitz}), we
see that while this has only a modest effect, it does somewhat suppress, about a 10\% reduction, the events in the dead zone.
Note that these plots are all made with the same set of parameters.

\section{Assessing the Logarithmic Accuracy}
\label{sec:Lund}

The angular-ordered parton shower has the correct single-emission probability by construction.
However it is still instructive to calculate the Lund variables,
\emph{i.e.}\ the transverse momentum $k_\perp$ and rapidity $y$, to see how
the \Herwig{} variables relate to the physical ones.
For a single gluon emission, $m_0=m_1=m$ and $m_2=0$, all three choices for the interpretation of the evolution
variable are identical, giving
\begin{subequations}
\begin{align}
  k^2_\perp &= p^2_T = (1-z)^2\left(z^2\tilde{q}^2-m^2\right)                                 & &\approx z^2(1-z)^2\tilde{q}^2 & &\approx \epsilon^2\tilde{q}^2, \\
  y      &= \frac12\ln \left[ \frac{(1 + b - c + \lambda)^2 Q^2 (1 - z)^2}{4p_T^2} \right] & &\approx\ln\left[\frac{Q(1-z)}{p_T}\right]
         & &\approx  \ln\left[\frac{Q}{\tilde{q}}\right],
\end{align}
\label{eqn:lund_single}
\end{subequations}
where $\lambda=\lambda(1,b,c)$. The first approximation is that both the
radiating particle and the spectator are massless, \emph{i.e.}\ $m\to0$,
and the second approximation is that the emitted gluon is soft, \emph{i.e.}
$z=1-\epsilon$ with $\epsilon\to0$.
The \Herwig{} soft collinear gluon emission probability from a massless quark line is given by 
\begin{equation}
d P_{\rm soft}^{\tt Hw7} = C_F \frac{{\rm d}\tilde{q}^2}{\tilde{q}^2}\frac{\alpha_S (z(1-z)\tilde{q})}{2\pi}
 {\rm d} z \frac{(1+z^2)}{1-z}\frac{{\rm d} \phi}{2\pi} \approx C_F \frac{{\rm d}\tilde{q}^2}{\tilde{q}^2}\frac{\alpha_S (\epsilon\tilde{q})}{\pi} \frac{{\rm d} \epsilon}{\epsilon}\frac{{\rm d} \phi}{2\pi},
\end{equation}
if we rearrange the above expression in terms of the Lund variables $k_T$ and $y$ 
we reproduce the correct form of the soft collinear emission probability
\begin{equation}
{\rm d} \mathcal{P} = C_{\rm F}\frac{\alpha_s(k_{\perp})}{\pi}\, \frac{d k_{\perp}^2}{k_{\perp}^2}\, d y\,\frac{d\phi}{2\pi}.
\end{equation}

We now need to investigate the accuracy for two successive gluon
emissions, \emph{i.e.}\ $m_{0,1,3}=m$, $m_{2,4}=0$. In particular, in
angular-ordered parton showers, one can obtain strongly disordered
regions in which a second emission is much harder (in energy,
contribution to jet virtuality or transverse momentum) than the
first. We therefore have to check that the kinematics of the softer
first gluon are not disturbed by the second harder one.

The different schemes only affect the relationship between the
transverse momenta and the evolution variable, this means that the kinematics are the same in all three schemes when expressed in terms of the transverse momenta.
The Lund variables for the two emissions are therefore:
\begin{subequations}
  \begin{align}
    k_{\perp1}^2 &= p^2_{T1};\\
    y_1        &= \frac12\ln\left[\frac{(1 + b - c + \lambda)^2 Q^2 (1 - z_1)^2}{4 p_{T1}^2}\right];\\
    k_{\perp2}^2 &= ({\bf p}_{T2}-(1-z_2){\bf p}_{T1})^2;\\
    y_2 &=\frac12\ln\left[\frac{(1 + b - c + \lambda)^2 Q^2 z_1^2 (1 - z_2)^2}
      {4  k_{\perp2}^2 }\right].
\end{align}\end{subequations}

All three choices of evolution variable are identical for one emission, therefore
\begin{equation}
p_{T2}^2 =   (1-z_2)^2\left[z_2^2\tilde{q}^2_2-m^2\right],
\end{equation}
and the virtuality of the branching parton is
\begin{equation}
  q^2_1 = z_2(1-z_2)\tilde{q}^2_2+m^2.
\end{equation}

For the first branching the relationships depend on our choice of reconstruction scheme.

\subsection[$p_T$ preserving scheme]{\boldmath{$p_T$} preserving scheme}

If we use the $p_T$ preserving scheme
\begin{equation}
p_{T1}^2 = (1-z_1)^2\left[z_1^2\tilde{q}_1^2-m^2\right],
\end{equation}
the final virtual mass of the original parton is
  \begin{equation}
    q_0^2 = \frac{p_{T1}^2}{z_1 (1 - z_1 )} +\frac{q_1^2}{z_1} =
    z_1 (1 - z_1 )\tilde{q}_1^2 +\frac{z_2 (1 - z_2 )\tilde{q}_2^2}{z_1}+m^2,
  \end{equation}
  and
  \begin{equation}
    p_{T2}^2 = (1-z_2)^2\left(z_2\sqrt{\tilde{q}^2_2-\frac{m^2}{z_2^2}}{\bf \hat{n}}_{2}-z_1(1-z_1) \sqrt{\tilde{q}^2_1-\frac{m^2}{z_1^2}} {\bf \hat{n}}_{1}\right)^2,
  \end{equation}
where we recall that ${\bf \hat{n}}_{i}$ is a unit vector parallel to ${\bf p}_{T i}$, see  Eqn.~\eqref{eqn:nhat}.
 
\begin{figure}[tb]
\centering
\includegraphics[width=0.75\textwidth]{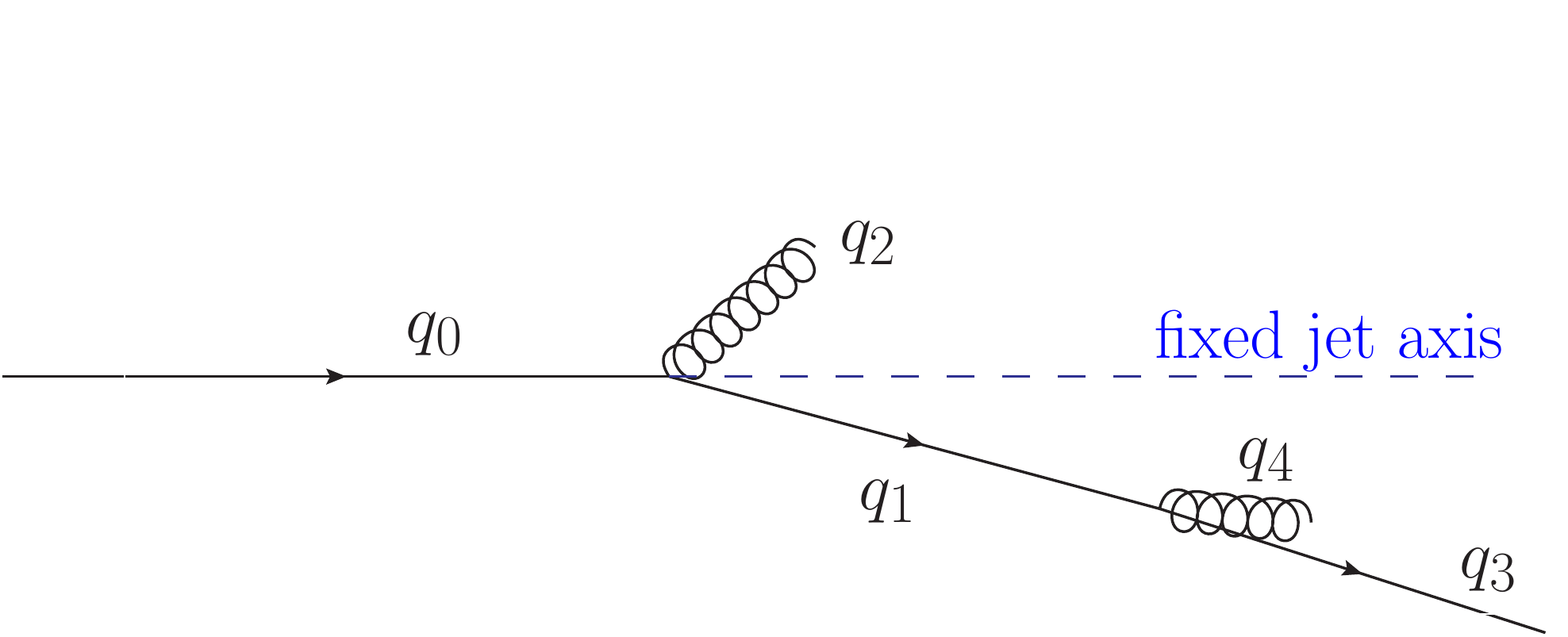}
\caption{Region in which the emission angle of the second
gluon is smaller than the recoil angle of the quark from the first-gluon emission.}
\label{fig:disorderedRegion}
\end{figure}

In the massless and soft limits, $z_{1,2}\to 1$ such that $z_{1,2}=1-\epsilon_{1,2}$ and $\epsilon_{1,2}\ll1$, the Lund variables are
\begin{subequations}
  \begin{align}
    k_{\perp1}^2 &\approx \epsilon_1^2\tilde{q}_1^2 ;\\
    y_1        &\approx\ln\left[\frac{Q}{\tilde{q}_1}\right]\\
    k_{\perp2}^2&\approx\epsilon_2^2(\tilde{q}_2{\bf \hat{n}}_{2}-\epsilon_1\tilde{q}_1{\bf \hat{n}}_{1})^2;\\
    y_2 &\approx\frac12\ln\left[\frac{Q^2}{(\tilde{q}_2{\bf \hat{n}}_{2}-\epsilon_1\tilde{q}_1{\bf \hat{n}}_{1})^2}  \right];
\end{align}\end{subequations}
In the soft limit
  \begin{equation}
    q_0^2 = \epsilon_1 \tilde{q}_1^2 + \epsilon_2 \tilde{q}_2^2+m^2.
  \end{equation}
  As the limit from angular-ordering is $\tilde{q}_1\geq\tilde{q}_2$ we see that for
  \begin{equation}
    \epsilon_2 \tilde{q}_2^2 > \epsilon_1 \tilde{q}_1^2,
  \end{equation}
  there is a disordered region where the contribution of a second harder gluon to the virtuality of
  the original parton is dominant. 
In this disordered region, \mbox{$k_{\perp2}\gg k_{\perp1}$} so that we can neglect $\epsilon_1\tilde{q}_1$
relative to $\tilde{q}_2$ and the kinematics are effectively independent.
However, there is a region in which the transverse momentum of the first emission overwhelms
that of the second, if \mbox{$\tilde{q}_2 < \epsilon_1 \tilde{q}_1= k_{\perp1}$}. This is the region in which the emission angle of the second
gluon is smaller than the recoil angle of the quark from the first gluon~(Fig.~\ref{fig:disorderedRegion}). It is an issue because we
have measured the transverse momentum and rapidity relative to the fixed jet axis, not the local
axis of emission\footnote{Similar issues were discussed in the context
of CAESAR resummation, see Ref.~\cite{Banfi:2004yd} Appendix~C.}.
If we calculate the Lund variables using $q_3$ as the axis:
\begin{subequations}
  \begin{align}
    k_{\perp1}^2 &\approx \epsilon^2_1( \tilde{q}_1 {\bf \hat n}_1+\epsilon_2\tilde{q}_2 {\bf \hat n}_2)^2;\\
    y_1        &\approx \frac12\ln\left[\frac{Q^2}{(\tilde{q}_1 {\bf \hat n}_1+\epsilon_2\tilde{q}_2 {\bf \hat n}_2)^2}\right];\\
    k_{\perp2}^2 &\approx \epsilon_2^2\tilde{q}_2^2;\\
    y_2 &\approx \ln\left[\frac{Q}{\tilde{q}_2}\right].
\end{align}\end{subequations}
The second gluon variables are now the same as the single emission case, Eqn.\,\ref{eqn:lund_single}, thus retaining the correct behaviour in the soft limit.
The first gluon variables are correct this time, because $\tilde{q}_2$ is
always smaller than $\tilde{q}_1$ and the factor of $\epsilon_2$ makes it arbitrarily smaller. Thus, this scheme is accurate to leading logarithmic order as it reproduces the correct behaviour of the soft, collinear splitting function.

\subsection[$q^2$ preserving scheme]{\boldmath{$q^2$} preserving scheme}

For the $q^2$ preserving scheme
  \begin{eqnarray}
    p_{T1}^2 &=& \max\left(z_1^2 (1 - z_1 )^2 \tilde{q}_1^2 +m^2z_1(1-z_1)- q_1^2(1-z_1),0\right)\nonumber\\ &=&
    \max\left((1-z_1)\left[(1-z_1)(z_1^2\tilde{q}^2_1-m^2) - z_2 (1 - z_2 )\tilde{q}_2^2 \right],0\right),
    \label{eqn:qtilde2_pT}
  \end{eqnarray}
  so that the transverse momentum is non-zero if
  \begin{equation}
    (1-z_1)(z_1^2\tilde{q}^2_1-m^2)  >z_2 (1 - z_2 )\tilde{q}_2^2.
\end{equation}
In the limit that both $z_{1,2}\to 1$ then
\begin{equation}
 p_{T1}^2 = \max\left( \epsilon_1 (\epsilon_1 (\tilde{q}_1^2-m^2) - \epsilon_2 \tilde{q}_2^2) ,0\right),
\end{equation}
so that in the soft limit the transverse momentum is non-zero for massless partons if
\begin{equation}
  \epsilon_1\tilde{q}^2_1 > \epsilon_2\tilde{q}^2_2, 
\end{equation}
which is effectively the requirement that the generated virtualities are ordered, which is clearly violated
in the disordered region we are concerned about.

In the ordered region in which a solution is possible, the Lund variables, calculated relative to the $q_3$ axis
are:
\begin{subequations}
  \begin{align}
    k_{\perp1}^2 &\approx \epsilon^2_1\tilde{q}^2_1-\epsilon_1\epsilon_2\tilde{q}^2_2; \\
    y_1        &\approx \frac12\ln\left[\frac{Q^2}{\tilde{q}^2_1-\tilde{q}^2_2\frac{\epsilon_2}{\epsilon_1}}\right];\\
    k_{\perp2}^2 &\approx \epsilon_2^2\tilde{q}_2^2;\\
    y_2 &\approx \ln\left[\frac{Q}{\tilde{q}_2}\right].\
\end{align}\label{eqn:lund2}\end{subequations}
In the bulk of the region, the $\tilde{q}^2_2$ terms are negligible. However, along
the ``line'' $\epsilon_2\tilde{q}^2_2\sim\epsilon_1\tilde{q}^2_1$ the generated $k_{\perp1}^2$ value
is wrong by a factor of order 1. Moreover, for most reasonable event shapes, {\it e.g.}\ thrust,
the first gluon is the dominant one. Therefore this is a next-to-leading-logarithmic~(NLL) error, \emph{i.e.}\
the double logarithmic behaviour is correct, while the single soft logarithm is incorrect.
An explicit derivation for the case of the thrust is given in Appendix~\ref{sec:thrustLogs}.
  
In the disordered region, $p_{T1}=0$, therefore the Lund variables are:
\begin{subequations}
  \begin{align}
    k_{\perp1}^2 &\approx \epsilon^2_1 p_{T2}^2 &&\approx \epsilon^2_1\epsilon^2_2 \tilde{q}^2_2; \\
    y_1        &\approx \frac12\ln\left[\frac{Q^2}{p_{T2}^2}\right] &&\approx \frac12\ln\left[\frac{Q^2}{\epsilon^2_2 \tilde{q}^2_2}\right];
  \end{align}\end{subequations}
with $k_{\perp2}^2$ and $y_2$ given by Eqn.\,\ref{eqn:lund2}.
While the kinematics of the second gluon are correct, kinematics of the first gluon are completely wrong in this
region in the Lund plane. This could, in principle, be a leading-log effect. However, for the example of the thrust distribution, 
in this region the second gluon is the hardest one and the first gluon
gives a sub-leading contribution to the observable.
Therefore, again, it is only along the line at the
edge of this region that one gets a significant effect and it is a NLL error.
We conclude that the $q^2$ preserving looks undesirable, in reconstructing
incorrect kinematics over a finite area of the Lund plane. In practice this leads to a NLL error in
the thrust distribution (see Appendix~\ref{sec:thrustLogs}).
Related problems with the $q^2$-preserving scheme were also noted in
Ref.~\cite{Hoang:2018zrp}.

\subsection{Dot-product preserving scheme}

In the dot-product preserving scheme the transverse momentum of the second branching is unchanged but for the first it becomes
  \begin{equation}
 p_{T1}^2 = z_1^2(1 - z_1 )^2 \tilde{q}_1^2 - q_1^2(1-z_1)^2 =  (1 - z_1 )^2 \left[z_1^2\tilde{q}_1^2 - z_2 (1 - z_2 )\tilde{q}_2^2 -m^2\right].
  \end{equation}
  The difference relative to Eqn.\,\ref{eqn:qtilde2_pT} looks minor, but now we have to compare $\tilde{q}^2_1$ with $\epsilon_2\tilde{q}^2_2$,
  $\tilde{q}^2_2$ has to be smaller than $\tilde{q}^2_1$ and the factor of $\epsilon_2$ makes it parametrically smaller. The second term can
  therefore never be as large as the first.

  The virtuality of the first parton is
  \begin{equation}
    q^2_0 = \tilde{q}^2_1z_1(1-z_1)+\tilde{q}^2_2z_2(1-z_2)+m^2,
  \end{equation}
  which for soft emissions can be dominated by the second emission for $\epsilon_2>\epsilon_1$.
  In this case the transverse momentum of the second branching is
  \begin{equation}
    p_{T2}^2 =  (1-z_2)^2\left(z_2\sqrt{\tilde{q}^2_2-\frac{m^2}{z_2^2}}{\bf \hat{n}}_{2}-z_1(1-z_1)\sqrt{\tilde{q}^2_1-\frac{m^2}{z_1^2}-\frac{z_2(1-z_2)\tilde{q}^2_2}{z_1^2}}{\bf \hat{n}}_{1}\right)^2 
  \end{equation}
  
In the massless and soft limits the Lund variables, with respect to the direction of $p$, are
\begin{subequations}\begin{align}
    k_{\perp1}^2 &\approx \epsilon_1^2(\tilde{q}_1^2-\epsilon_2 \tilde{q}_2^2) ;\\
    y_1        &\approx \frac{1}{2}\ln\left[\frac{Q^2}{\tilde{q}_1^2-\epsilon_2 \tilde{q}_2^2}\right]\\
    k_{\perp2}^2 &\approx\epsilon_2^2(\tilde{q}_2{\bf \hat{n}}_{2}-\epsilon_1\tilde{q}_1{\bf \hat{n}}_{1})^2;\\
    y_2 &\approx\frac12\ln\left[\frac{Q^2} {\left(\tilde{q}_2{\bf \hat{n}}_{2}-\epsilon_1\tilde{q}_1{\bf \hat{n}}_{1}\right)^2  }\right],
\end{align}\end{subequations}
while with respect to the direction of $q_3$ they become
\begin{subequations}
  \begin{align}
    k_{\perp1}^2 &\approx \epsilon_1^2(\tilde{q}_1^2+\epsilon_2 \tilde{q}_2^2) ;\\
    y_1        &\approx \frac{1}{2}\ln\left[\frac{Q^2}{\tilde{q}_1^2+\epsilon_2 \tilde{q}_2^2}\right]\\
    k_{\perp2}^2 &\approx\epsilon_2^2\tilde{q}_2^2;\\
    y_2 &\approx\ln\left[\frac{Q} {\tilde{q}_2}\right].
\end{align}\end{subequations}
 
\subsection{Global recoil}

  We also need to consider the impact of the implementation of the
global recoil in \textsf{Herwig 7}. For simplicity we will consider the case of two final-state particles,
the generic case can be found in Ref.~\cite{Bahr:2008pv}.
We have a particle $a$ with momentum
\begin{equation}
q_a = \sqrt{s}\left[ 1,0,0,0 \right],
\end{equation}
which splits into particles $b$ and $c$, whose momenta are given by
\begin{align}
p_b &= \frac{\sqrt{s}}{2} \left[ 1+b-c, 0, 0, +{\lambda\left( 1,b,c \right)} \right], &
p_c &= \frac{\sqrt{s}}{2} \left[ 1-b+c, 0, 0,-{\lambda\left( 1,b,c\right)} \right], 
\end{align}
where $\lambda$ is the K\"{a}ll\'{e}n function defined in Eqn.~\eqref{eqn:Kallen} and $b=m^2_b/s$, $c=m^2_c/s$. 
During the shower evolution the particles acquire a virtuality $q^2_b= b^\prime s$ and $q_c^2=c^\prime s$ and their momenta are modified 
\begin{align}
q_b &= p_b + \beta_b \, n_b, \\
q_c &= p_c + \beta_c \, n_c, 
\end{align}
where
\begin{align}
n_b &= \frac{\sqrt{s}}{2}{\lambda\left( 1,b,c\right)} \left[ 1, 0, 0, -1 \right], &
n_c &= \frac{\sqrt{s}}{2}{\lambda\left( 1,b,c\right)} \left[ 1, 0, 0, +1 \right],
\end{align}
and
\begin{equation}
\beta_b = \frac{s(b^\prime -b)}{2 p_b \cdot n_b}\,, \qquad \beta_c = \frac{s(c^\prime -c)}{2 p_c \cdot n_c}.
\end{equation}
However, if we want to have two particles with invariant mass $q^2_b$ and $q^2_c$, whose three-momentum is parallel to the direction of $p_b$ and $p_c$ respectively, the two particles must have four-momentum equal to
\begin{align}
q^\prime_b &= \frac{\sqrt{s}}{2} \left[ 1+b^\prime-c^\prime, 0, 0, +{\lambda\left( 1,b^\prime,c^\prime \right)} \right], &
q^\prime_c &= \frac{\sqrt{s}}{2} \left[ 1-b^\prime+c^\prime, 0, 0,-{\lambda\left( 1,b^\prime,c^\prime\right)} \right].
\end{align}
As $q_b+q_c=q^\prime_b+q^\prime_c$, they can be simply related by a Lorentz transform along the $p_b$~($p_c$) direction. 
The boost parameter for $b$ is
{\small\begin{equation}
\beta^{(b)} = \frac{((b+b^\prime) (1+b-c)+\lambda (b-b^\prime)) ((b-b^\prime) (1+b-c)+\lambda (b+b^\prime))-4 b^2 \lambda^\prime (1+b^\prime-c^\prime)}{((b-b^\prime) (1+b-c)+\lambda (b+b^\prime))^2+4 b^2 (1+b^\prime-c^\prime)^2}\,,
\end{equation}}%
where we have used the shorthand notation $\lambda = \lambda(1,b,c)$ and $\lambda^\prime=\lambda(1,b^\prime, c^\prime)$.
The expression may look complicated, but if we consider that $b$, $c$, $b^\prime$ and $c^\prime$ are all much smaller than 1, we get
\begin{align}
\beta^{(b)} &\approx c^\prime -c,   &  \beta^{(c)} &\approx b^\prime -b.
\end{align}
Also the partons which have $q_b$~($q_c$) as shower progenitor need to be boosted along the direction of the progenitor. This boost will leave the transverse momentum, the light-cone momentum $z$ and the ordering variable $\tilde{q}$ (since it is expressed in terms of scalar products and $z$) invariant, but not the rapidity of the particles.

Indeed the rapidities of partons having the $b$ as shower progenitor are slightly shift towards smaller values
\begin{equation}
\Delta y_b = \frac{1}{2} \log \left( \frac{1-\beta^{(b)}}{1+\beta^{(b)}} \right) \approx -\beta^{(b)},
\end{equation} 
and the rapidities of those coming from the $c$ cascade are slightly pulled in the opposite direction
\begin{equation}
\Delta y_c = \frac{1}{2} \log \left( \frac{1+\beta^{(c)}}{1-\beta^{(c)}} \right) \approx \beta^{(c)},
\end{equation} 
where we expand the result because the boost parameter is generally much smaller than $1$, being of the order of $(q^2-m^2)/s$, where $q^2$ is the virtuality developed by the colour partner of the shower progenitor and $m^2$ its mass.

Let us now discuss the impact of global recoil for soft emission in the
massless limit, \emph{i.e.}\ for $b=c=0$. Let us assume for simplicity that $b$ is a
quark $q$ and $c$ is an anti-quark $\bar{q}$.
If we use the default \textsf{Herwig 7} settings, partons originated from $b$ will all have positive rapidity and the single emission probability in the soft limit is
\begin{equation}
d P_{q \to qg} =  C_{\rm F}\frac{\alpha_s(p_T)}{\pi}\,\frac{d\phi}{2\pi}\, \frac{d p_T^2}{p_T^2}\, d y \,\Theta(y)\,,
\end{equation} 
while the probability of a soft-emission originated from $c$ is given by
\begin{equation}
d P_{\bar{q} \to \bar{q}g} =  C_{\rm F}\frac{\alpha_s(p_T)}{\pi}\,\frac{d\phi}{2\pi}\, \frac{d p_T^2}{p_T^2}\, d y \,\Theta(-y)\,,
\end{equation} 
and the sum of the two contributions yields
\begin{equation}
d P_{\rm soft} = C_{\rm F}\frac{\alpha_s(p_{T})}{\pi}\,\frac{d\phi}{2\pi}\, \frac{d p_{T}}{p_{T}^2}\, d y.
\label{eqn:softEmissionProb}
\end{equation}
However, after we apply our global recoil, the rapidity of the partons gets
shifted, to the left for partons coming from $b$ and to the right for those coming from $c$, causing a double counting of the central-rapidity region.
If we call $\bar{\beta}$ the average boost-parameter that is applied after the global recoil, Eqn.~\eqref{eqn:softEmissionProb} will be modified to
\begin{equation}
d P_{\rm soft}^{\tt Hw7} =  C_{\rm F}\frac{\alpha_s(p_T)}{\pi}\,\frac{d\phi}{2\pi}\, \frac{d p_T^2}{p_T^2}\, d y \left[ 1+\Theta\left(|y|<\bar{\beta}\right)\right].
\end{equation} 
Nevertheless, given the fact that $\bar{\beta}$ is of the order $q^2/s$ and for soft emission typically $q^2 \ll s$, this is a power-suppressed effect, \emph{i.e.}\ non-logarithmic,
and therefore does not alter the logarithmic accuracy of the parton shower.

\section{Tuning}
\label{sec:tuning}

The new interpretation of the evolution variable means that the hadronization
parameters (which are highly sensitive to the PS algorithm) have to be
retuned. In order to do so, we follow the same strategy as in
Ref.~\cite{Reichelt:2017hts}: simulated events are analysed with
\textsf{Rivet}~\cite{Buckley:2010ar}, which also enables a comparison with
experimental results. The dependence on the hadronization and parton shower
parameters~\cite{hw7manual} is interpolated by the \textsf{Professor}
program~\cite{Buckley:2009bj}, which also finds the set of values which best
fit the experimental measurements.
In our case, where observables were measured
by multiple experiments, only the most recent set of data is used.  We have
not included LHC data in the tuning due to the high CPU-time requirement. We
consider only the transverse momentum~(pTmin) and not the virtuality as a cutoff parameter.

In order to tune the shower and light quark hadronization parameters we used
data on jet rates and event shapes for centre-of-mass energies between 14 and 44\,GeV \cite{Braunschweig:1990yd,MovillaFernandez:1997fr,Pfeifenschneider:1999rz,Achard:2004sv}, 
at LEP1 and SLD~\cite{Abreu:1996na,Barate:1996fi,Pfeifenschneider:1999rz,Abbiendi:2004qz,Heister:2003aj,Achard:2004sv} and LEP2~\cite{Pfeifenschneider:1999rz,Heister:2003aj,Abbiendi:2004qz,Achard:2004sv}, 
particle multiplicities~\cite{Abreu:1996na,Barate:1996fi} and spectra~\cite{Akers:1994ez,Alexander:1995gq,Alexander:1995qb,Abreu:1995qx,Alexander:1996qj,Abreu:1996na,Barate:1996fi,Ackerstaff:1997kj,Abreu:1998nn,Ackerstaff:1998ap,Ackerstaff:1998ue,Abbiendi:2000cv,Heister:2001kp,Acton:1992us,Adriani:1992hd} at LEP 1, identified particle spectra
below the $\Upsilon(4S)$ from Babar~\cite{Lees:2013rqd}, the charged particle multiplicity and distributions from ~\cite{Berger:1980zb,Derrick:1986jx,Aihara:1986mv,Bartel:1983qp,Braunschweig:1989bp,Zheng:1990iq}  for centre-of-mass energies between 14 and 61\,GeV,
the charged particle multiplicity~\cite{Ackerstaff:1998hz,Abe:1996zi} and particle spectra~\cite{Ackerstaff:1998hz,Abe:1998zs,Abe:2003iy} in light quark events at LEP1 and SLD,
the charged particle multiplicity in light quark events at LEP2~\cite{Abreu:2000nt,Abbiendi:2002vn}, the charged particle multiplicity distribution at LEP 1\cite{Decamp:1991uz,Abreu:1990cc,Acton:1991aa},
and hadron multiplicities at the Z-pole \cite{Amsler:2008zzb}, and data on the properties of gluon 
jets~\cite{Abbiendi:2003gh,Abbiendi:2004pr}.

The hadronization parameters for charm quarks were tuned using
the charged multiplicity in charm events at HRS~\cite{Aihara:1986mv}, SLD~\cite{Abe:1996zi} and
LEP2~\cite{Abreu:2000nt,Abbiendi:2002vn}, the light hadron spectra in charm events at LEP1 and SLD \cite{Ackerstaff:1998hz,Abe:1998zs,Abe:2003iy},
the multiplicities of charm hadrons at the Z-pole \cite{Abreu:1996na,Amsler:2008zzb}, and
charm hadron spectra below the $\Upsilon(4S)$~\cite{Seuster:2005tr,Aubert:2006cp,Artuso:2004pj} and at LEP1~\cite{Barate:1999bg}.

The hadronization parameters for bottom quarks were tuned using
the charged multiplicity in bottom events at HRS~\cite{Aihara:1986mv}, SLD~\cite{Abe:1996zi} and
LEP2~\cite{Abreu:2000nt,Abbiendi:2002vn,Abreu:2000nt}, the light hadron spectra and event shapes in bottom events at LEP1 and SLD \cite{Ackerstaff:1998hz,Abe:1998zs,Abe:2003iy,Abbiendi:2004pr,Achard:2004sv},
the multiplicities of charm and bottom hadrons at the Z-pole \cite{Abreu:1996na,Amsler:2008zzb},
charm hadron spectra at LEP1~\cite{Barate:1999bg,Alexander:1995qb} and the
bottom fragmentation function measured at LEP1 and SLD~\cite{Abe:2002iq,Heister:2001jg,DELPHI:2011aa,Abbiendi:2002vt}.

\textsf{Professor} offers the possibility to weight each observable differently: we adopted the same weights as in Ref.~\cite{Reichelt:2017hts}. Furthermore, as in \cite{Reichelt:2017hts}, to prevent the fit being dominated by a few observables with very small experimental uncertainty, we impose a minimum relative error of $5\%$ in the computation of the chi-squared $\chi^2$.

The following procedure is adopted to tune \textsf{Herwig\ 7}.
\begin{enumerate}
\item First the strong coupling computed in the CMW scheme~\cite{Catani:1990rr} $\alpha_s^{\rm CMW}$, the minimum transverse momentum allowed in the showering phase $p_{T}^{\text{min}}$, and the light quark hadronization parameters
are tuned to event shapes, charged-particle multiplicity and identified-particle spectra and rates which only involve light quark hadrons. This class of observables is labelled as ``general'' in Tab.~\ref{tab:chi2}.
\item The hadronization parameters for bottom quarks are then tuned to
the bottom quark fragmentation function, event shapes and to the identified-particle spectra from $b\bar{b}$ events.
\item The hadronization parameters involving charm quarks are then tuned
to identified-particle spectra and measurements of event shapes from charm events.\footnote{Charm parameters are the last to be determined, since charm hadrons are also produced from $b$-hadron decays.}
\item We then vary one parameter at a time to see if our tune corresponds to the minimum of the $\chi^2$. In case any of the parameters are significantly displaced from the minimum, we retune them all (this time considering all the experimental distributions for light, bottom and charm quarks together). 
\item We repeat the previous step except that now if any parameters are too far from the minimum of the $\chi^2$, their values are adjusted by hand. In particular, this is needed for bottom quark hadronization parameters like \textsf{ClMaxBottom} which \textsf{Professor} is not able to tune: this behaviour was also found in  Ref.~\cite{Reichelt:2017hts}.
\end{enumerate}

The values of the default parameters and the new ones we find with our tuning
procedure are shown in Tab.~\ref{tab:parameters}. The $\chi^2$ per degree of
freedom computed with the observables used for the tune, together with some recent data from the
ATLAS experiment~\cite{Aad:2016oit} which is sensitive to both quark and
gluon jet properties, are shown in
Tab.~\ref{tab:chi2}.

From
  Tab.~\ref{tab:parameters} we can notice that the four reconstruction
  choices correspond to four significantly different values of the strong
  coupling, where smaller values correspond to the schemes that give a poorer description of the non-logarithmically enhanced region of the spectrum. The
  introduction of the veto procedure in the dot preserving scheme indeed
  induces a $4\% $ enhancement in $\alpha_s$.

\begin{table*}
\resizebox{\textwidth}{!}
{
  \begin{tabular}{|c|C{1.7cm}|C{1.6cm}|C{1.5cm}|C{1.5cm}|C{1.5cm}|C{1.9cm}|}
\hline
    Preserved  & $p_T$ in \cite{Reichelt:2017hts}  & $q^2$ in \cite{Reichelt:2017hts} & $p_T$ &$q^2$ & $q_i \cdot q_j$ & $q_i \cdot q_j$+veto \\
\hline
\multicolumn{7}{|c|}{ \bf Light-quark hadronization and shower parameters} \\
\hline
AlphaMZ ($\alpha^{\rm CMW}_s(M_Z)$)      & 0.1087 & 0.1262 & 0.1074 & 0.1244 & 0.1136 & 0.1186\\
pTmin         & 0.933 & 1.223 & 0.900 & 1.136 & 0.924 & 0.958\\
ClMaxLight    & 3.639 & 3.003 & 4.204 & 3.141 & 3.653 & 3.649\\
ClPowLight    & 2.575 & 1.424 & 3.000 & 1.353 & 2.000 & 2.780\\
PSplitLight   & 1.016 & 0.848  & 0.914 & 0.831 & 0.935 & 0.899\\
PwtSquark     & 0.597 & 0.666  & 0.647 & 0.737 & 0.650 & 0.700\\
PwtDIquark    & 0.344 & 0.439 & 0.236 & 0.383 & 0.306 & 0.298 \\
\hline
\multicolumn{7}{|c|}{ \bf Bottom hadronization parameters} \\
\hline
ClMaxBottom             & 4.655 & 3.911 & 5.757 & 2.900 & 6.000 & 3.757\\
ClPowBottom             & 0.622 & 0.638 & 0.672 & 0.518 & 0.680 & 0.547\\
PSplitBottom            & 0.499 & 0.531 & 0.557 & 0.365 & 0.550 & 0.625\\
ClSmrBottom             & 0.082 & 0.020 & 0.117 & 0.070 & 0.105 & 0.078\\
SingleHadronLimitBottom & 0.000 & 0.000 & 0.000 & 0.000 & 0.000 & 0.000\\
\hline
\multicolumn{7}{|c|}{ \bf Charm hadronization parameters} \\
\hline
ClMaxCharm              & 3.551 & 3.638 & 4.204 & 3.564 & 3.796 & 3.950\\
ClPowCharm              & 1.923 & 2.332 & 3.000 & 2.089 & 2.235 & 2.559\\
PSplitCharm             & 1.260 & 1.234 & 1.060 & 0.928 & 0.990 & 0.994\\
ClSmrCharm              & 0.000 & 0.000 & 0.098 & 0.141 & 0.139 & 0.163\\
SingleHadronLimitCharm  & 0.000 & 0.000 & 0.000 & 0.011 & 0.000 & 0.000\\
\hline
  \end{tabular}
 }
  \caption{The Monte Carlo parameters obtained for different choices of the preserved quantity in the angular-ordered
  shower.}
  \label{tab:parameters}
\end{table*}

\begin{table*}
\centering
%\resizebox{\textwidth}{!}
%% {
   \begin{tabular}{|c||C{2.5cm}|C{2.5cm}|C{2.5cm}|C{2.5cm}|}
 \hline
     Preserved   &$p_T$  &  $q^2$ &
     $q_i \cdot q_j$ & $q_i \cdot q_j$+veto \\
 \hline
 \multicolumn{5}{|c|}{ \bf $\mathbf{\boldsymbol{\chi}^2}$ per d.o.f considering several set of observables} \\
 \hline
general & 4.406 &  3.152 & 3.735 & 3.352\\
\hline
bottom &5.964 &  6.494 & 5.127 & 4.118\\
\hline
charm  & 2.306 & 1.725 & 1.838 & 1.912\\
\hline
ATLAS jets &0.1598 & 0.4124 & 0.1925 & 0.5396\\
\hline
 \hline
 \multicolumn{5}{|c|}{ \bf $\mathbf{\boldsymbol{\chi}^2}$ per d.o.f
   considering sub-samples of the ``general'' observables } \\
 \hline
 mult & 3.031 & 2.757 & 2.822 & 2.776\\
 \hline
 event & 6.959&  3.461 & 5.191 & 3.877\\
 \hline
 ident & 10.706 & 9.950 & 9.777 & 10.105\\
 \hline 
jet & 4.579 & 3.226 & 4.093 & 3.638\\
\hline
gluon & 1.128 & 1.174 & 1.237 & 1.216\\
 \hline
charged & 5.439 & 2.515 & 3.724 & 2.856\\
\hline
   \end{tabular}
%%  }
  \caption{The $\chi^2$ per degree of freedom for different choices of the preserved quantity in the angular-ordered
  shower, obtained with the distributions we used to tune the light, bottom and
  charm parameters respectively. The $\chi^2$ corresponding to ATLAS jets, particle multiplicities~(mult), event shapes~(event),
  identified-particle spectra~(ident), quark jets~(jet), gluon
  jets~(gluon) and
  charged particle distributions~(charged) are also shown. }
  \label{tab:chi2}
\end{table*}

\section{Results}

In this section we present the results of our simulations, in order to compare the predictions obtained with the several implementations of the recoil discussed above. 
We first discuss the LEP results, for which \Herwig{} provides matrix-element corrections~(MEC), and then LHC ones for which \Herwig{} does not.

\subsection{LEP results}

\begin{figure}[tb]
\includegraphics[width=0.45\textwidth]{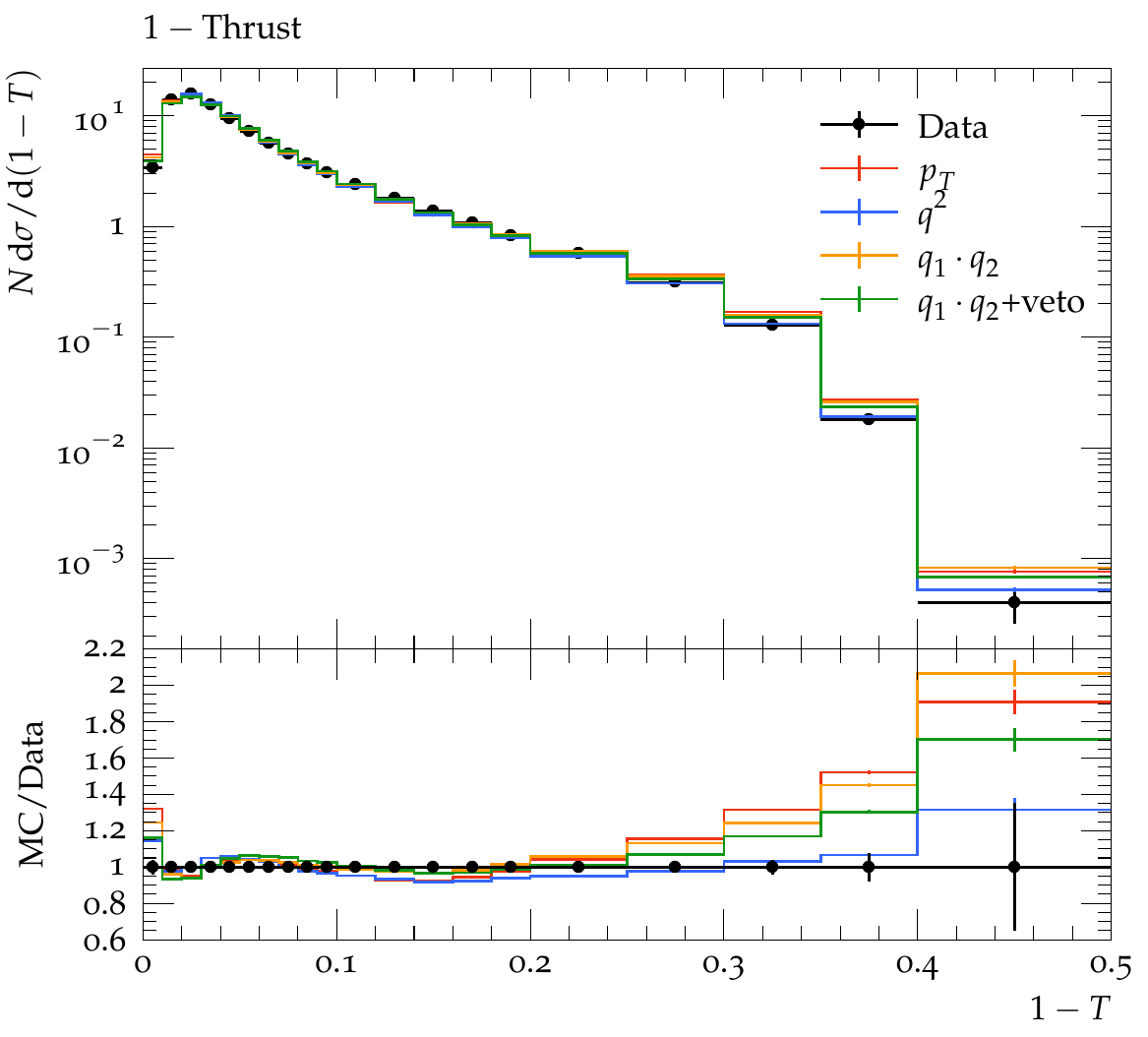}
\includegraphics[width=0.45\textwidth]{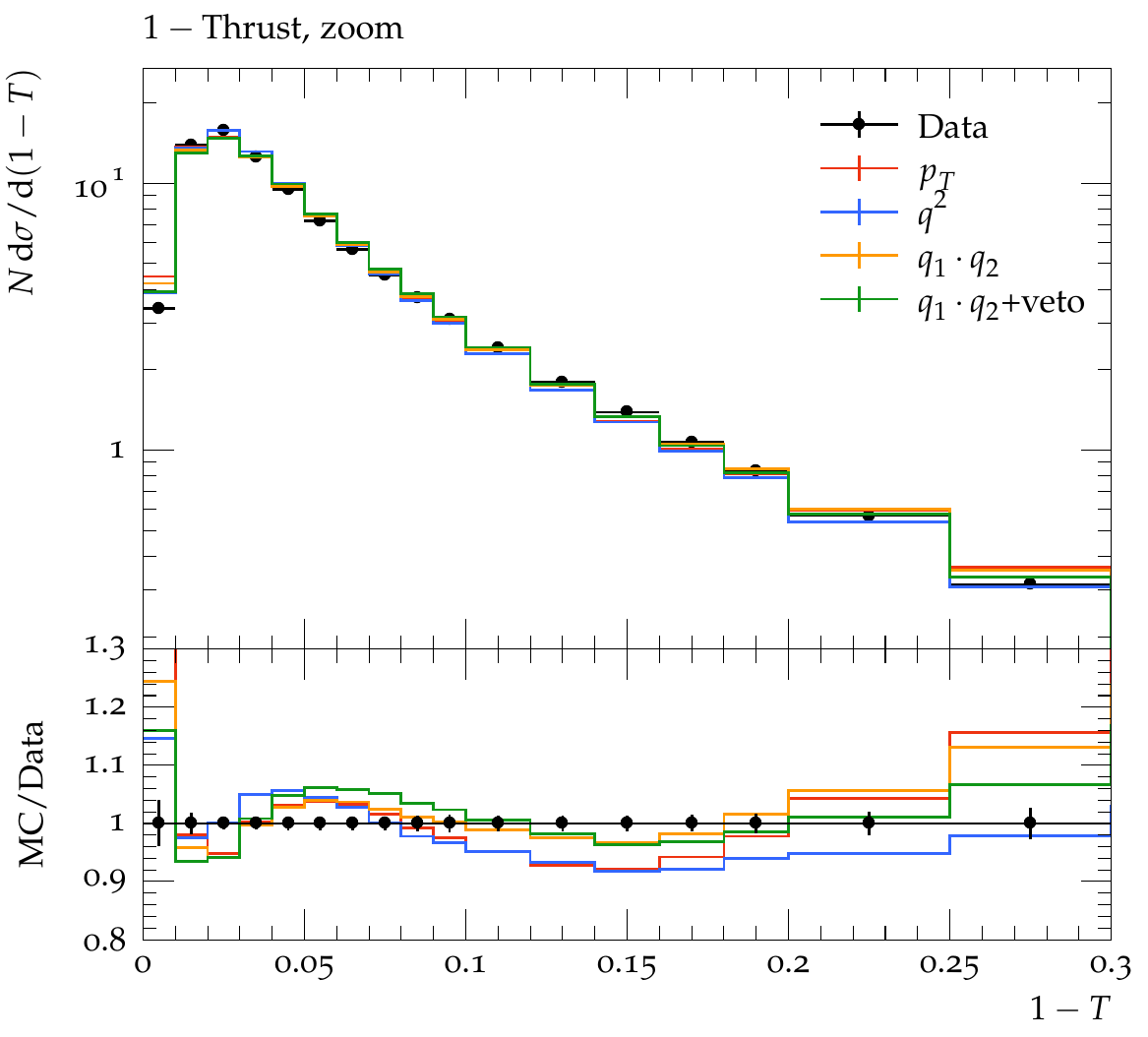}
\caption{The thrust at the Z-pole compared with data from the DELPHI~\cite{Abreu:1996na} experiment. In the right panel a zoom for small $1-T$ values is shown.}
\label{fig:thrust}
\end{figure}

The first event-shape distribution we consider is thrust,
Fig.~\ref{fig:thrust}. We find the well-known behaviour of the $p_T$-preserving
scheme, which overpopulates the non-loga\-rithmically-enhanced region of phase
space that is already filled by MEC and corresponds to the tail of the
distribution. Although the dot-product scheme performs better than the $p_T$
one it still overpopulates the \emph{dead zone}, however the description of the tail of the spectrum improves if we include the rejection veto described in Sec.~\ref{sec:veto}. In the right panel of Fig.~\ref{fig:thrust} an expanded view of the small $1-T$ region is displayed, where we notice that the new choice of the recoil yields a better agreement with data.

\begin{figure}[tb]
\includegraphics[width=0.45\textwidth]{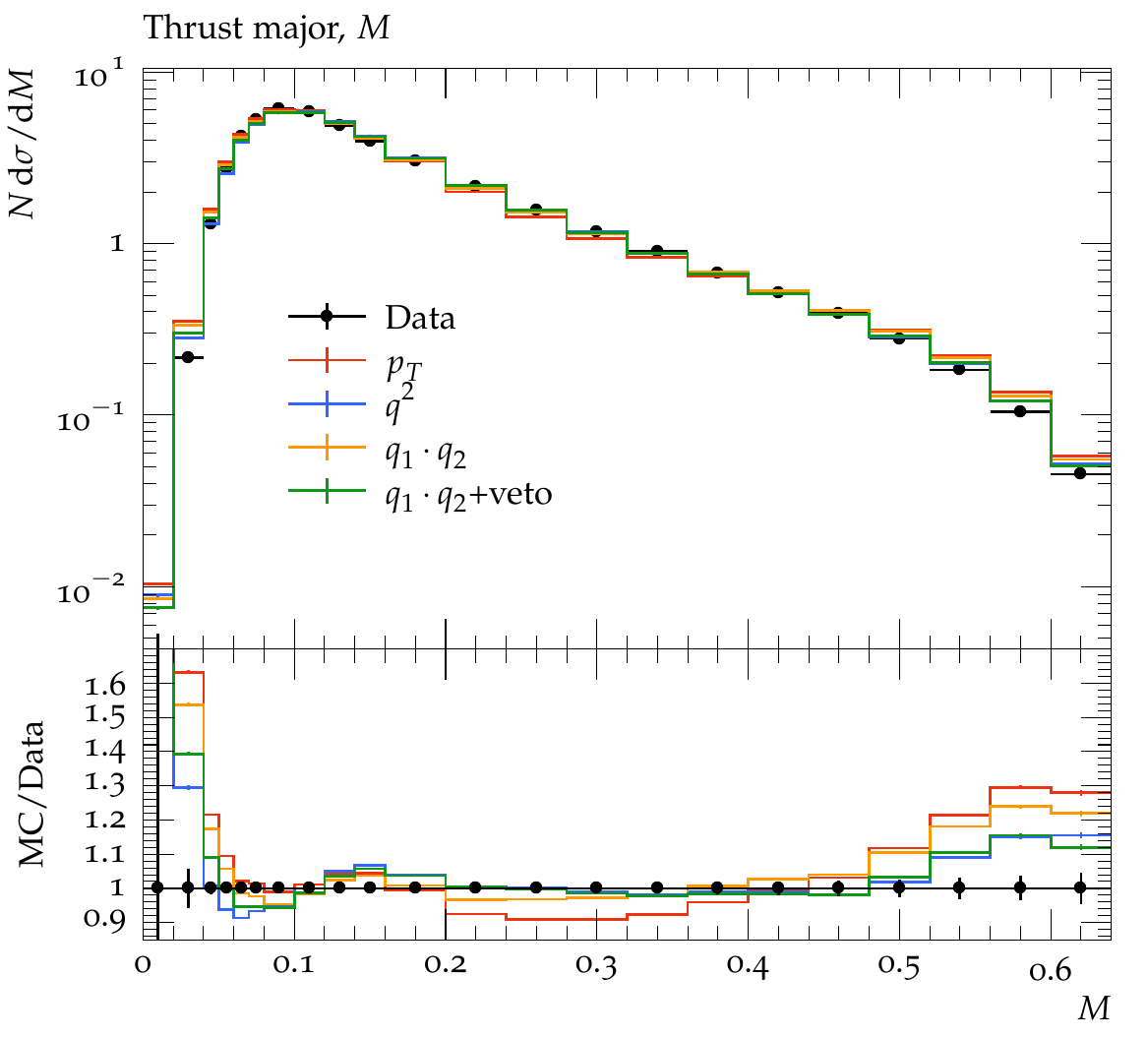}
\includegraphics[width=0.45\textwidth]{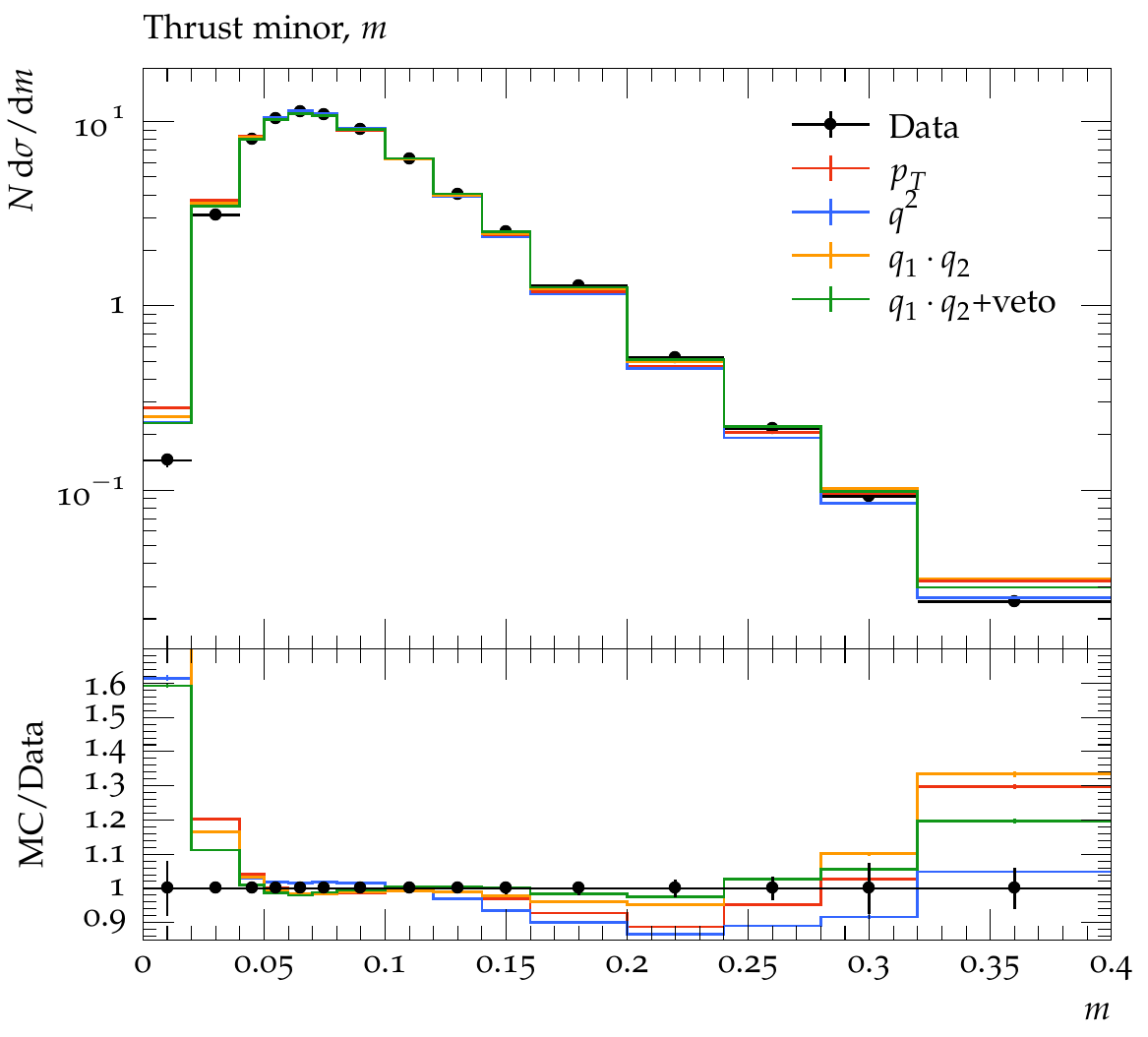}
\caption{Thrust major~(left) and minor~(right) at the Z-pole compared with data from the DELPHI~\cite{Abreu:1996na} experiment.}
\label{fig:thrust-Mm}
\end{figure}

\begin{figure}[tb]
\includegraphics[width=0.45\textwidth]{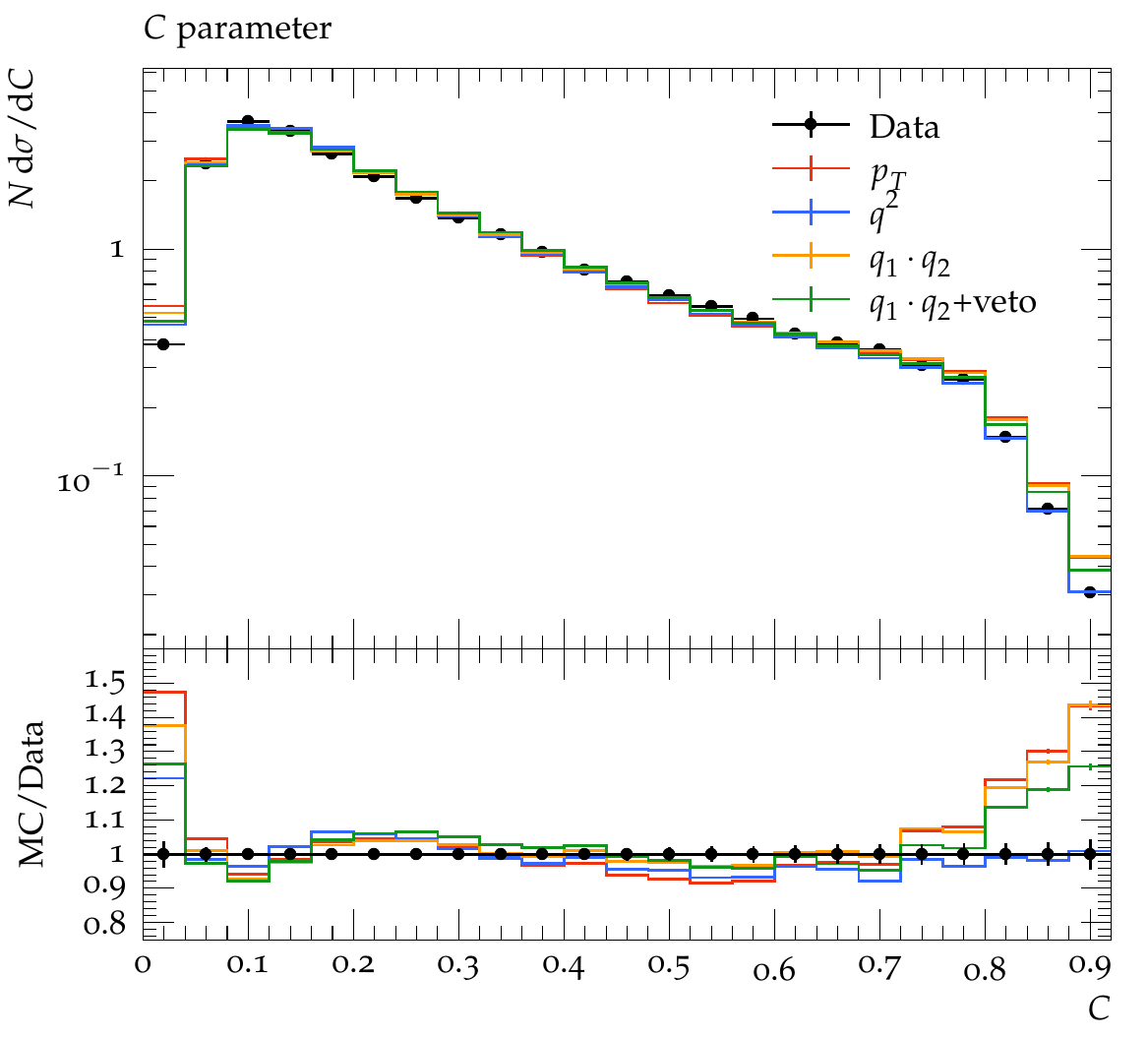}
\includegraphics[width=0.45\textwidth]{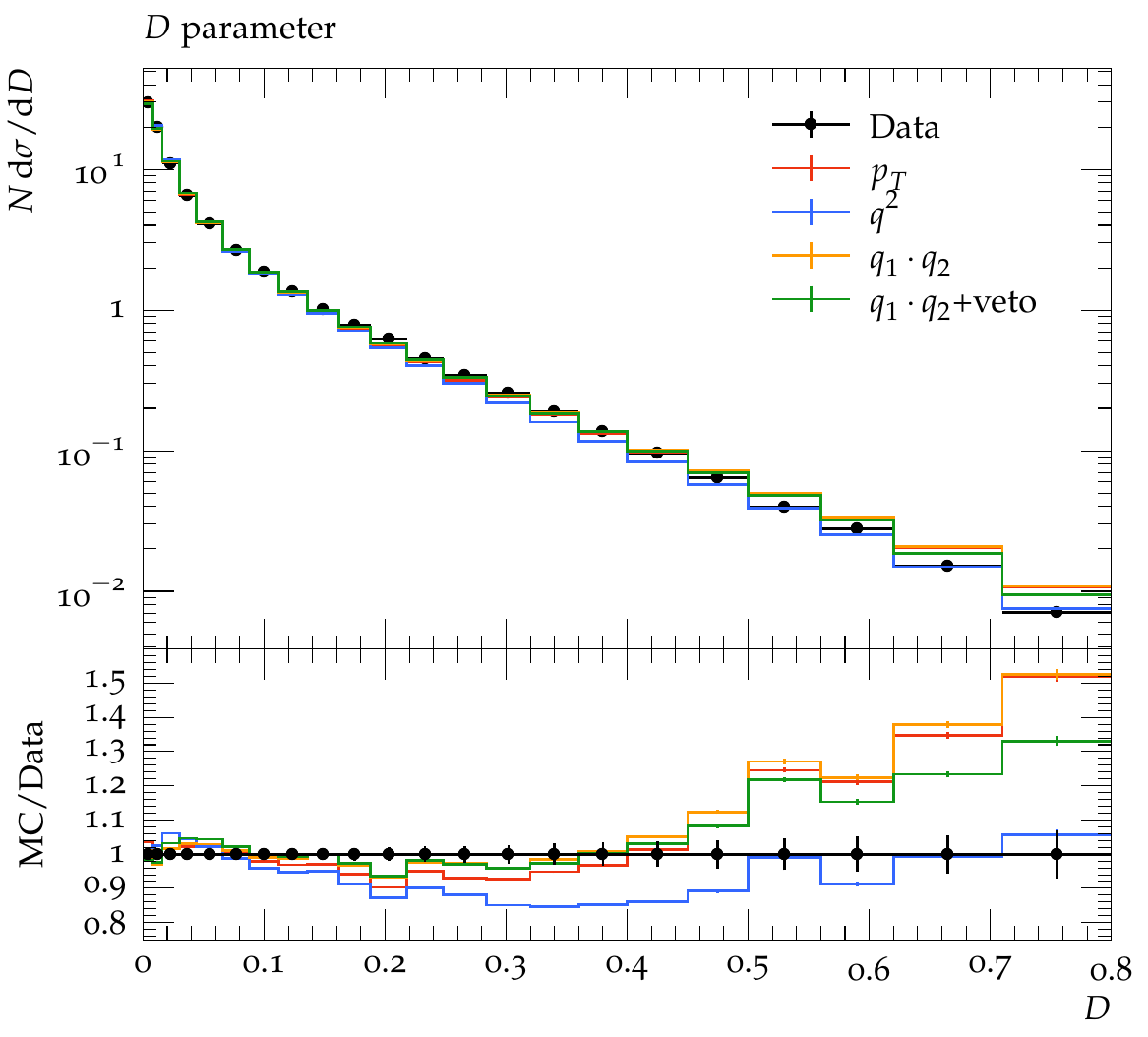}
\caption{C~(left) and D~(right) parameters at the Z-pole compared with data from the DELPHI~\cite{Abreu:1996na} experiment.}
\label{fig:CD}
\end{figure}

Very similar conclusions can be drawn from the thrust major and minor~(Fig.~\ref{fig:thrust-Mm}) distributions, and from the plots of the $C$- and $D$-parameters~(Fig.~\ref{fig:CD}).
For all the event shape distributions except for $D$, all the options
over-populate the first bin, but the $q^2$ and dot-product-plus-veto are
similar to each other and closest to the data.

\begin{figure}
\includegraphics[width=0.45\textwidth]{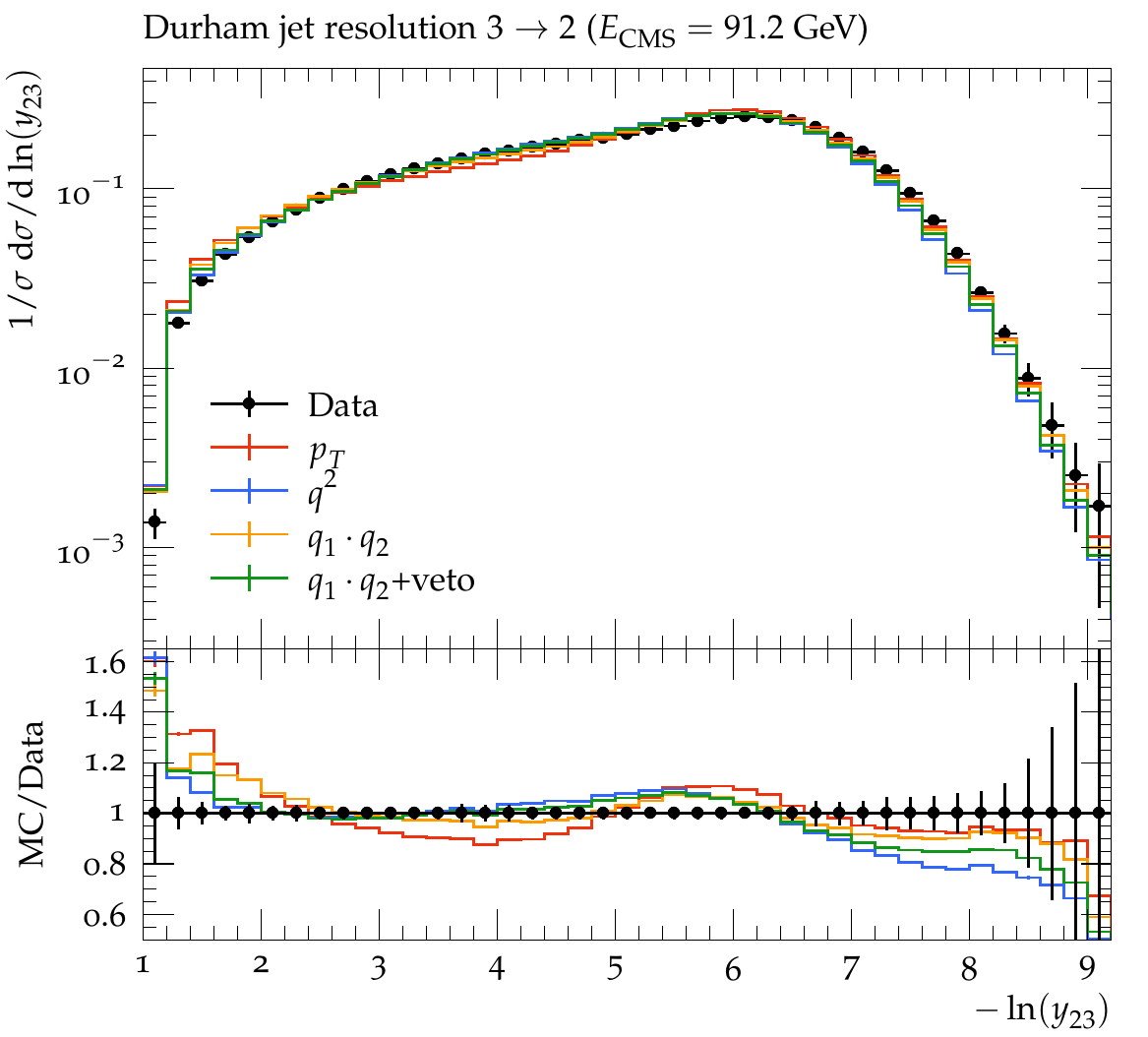}
\includegraphics[width=0.45\textwidth]{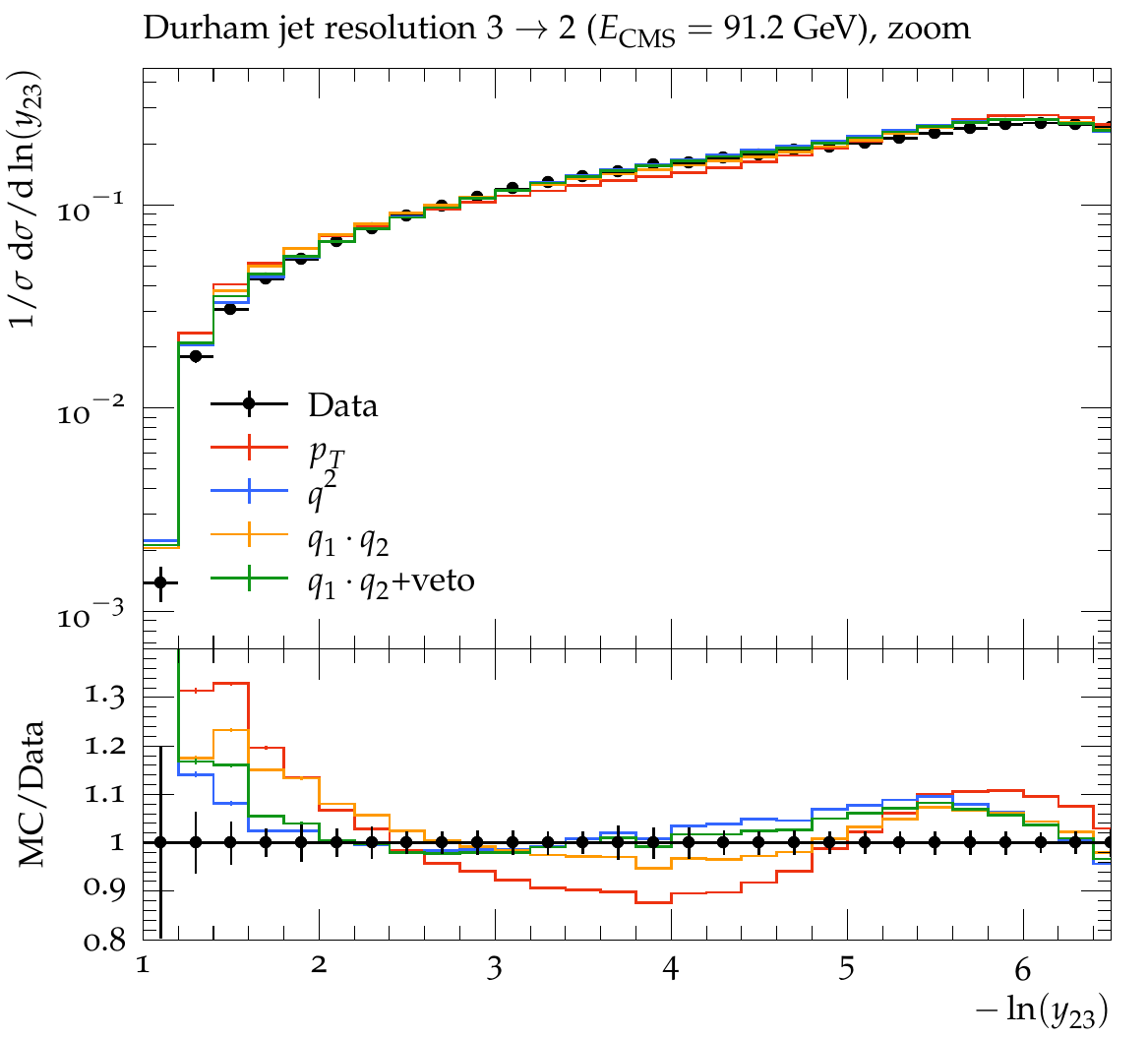}
\caption{Jet resolution parameter from a 3-jet configuration to a 2-jet configuration at the Z-pole compared with data from the ALEPH~\cite{Heister:2003aj} experiment. In the right panel an expanded section of the same plot is shown.}
\label{fig:y}
\end{figure}

Looking at the behaviour of the jet resolution parameter in Fig.~\ref{fig:y} we observe that the $p_T$-scheme most closely matches the data in the large $-\log(y_{23})$ (small $y_{23}$) tail of the distribution. However, in the small $-\log(y_{23})$ region the $q^2$ scheme yields a better description of the data. The dot-product scheme with the veto behaves very similar to the $q^2$ scheme, while the scheme without the veto is similar to the $p_T$ scheme in the tail of the distribution and to the $q^2$ one in the opposite limit, thus retaining the best description of the data over the whole range.

\begin{figure}
\includegraphics[width=0.45\textwidth]{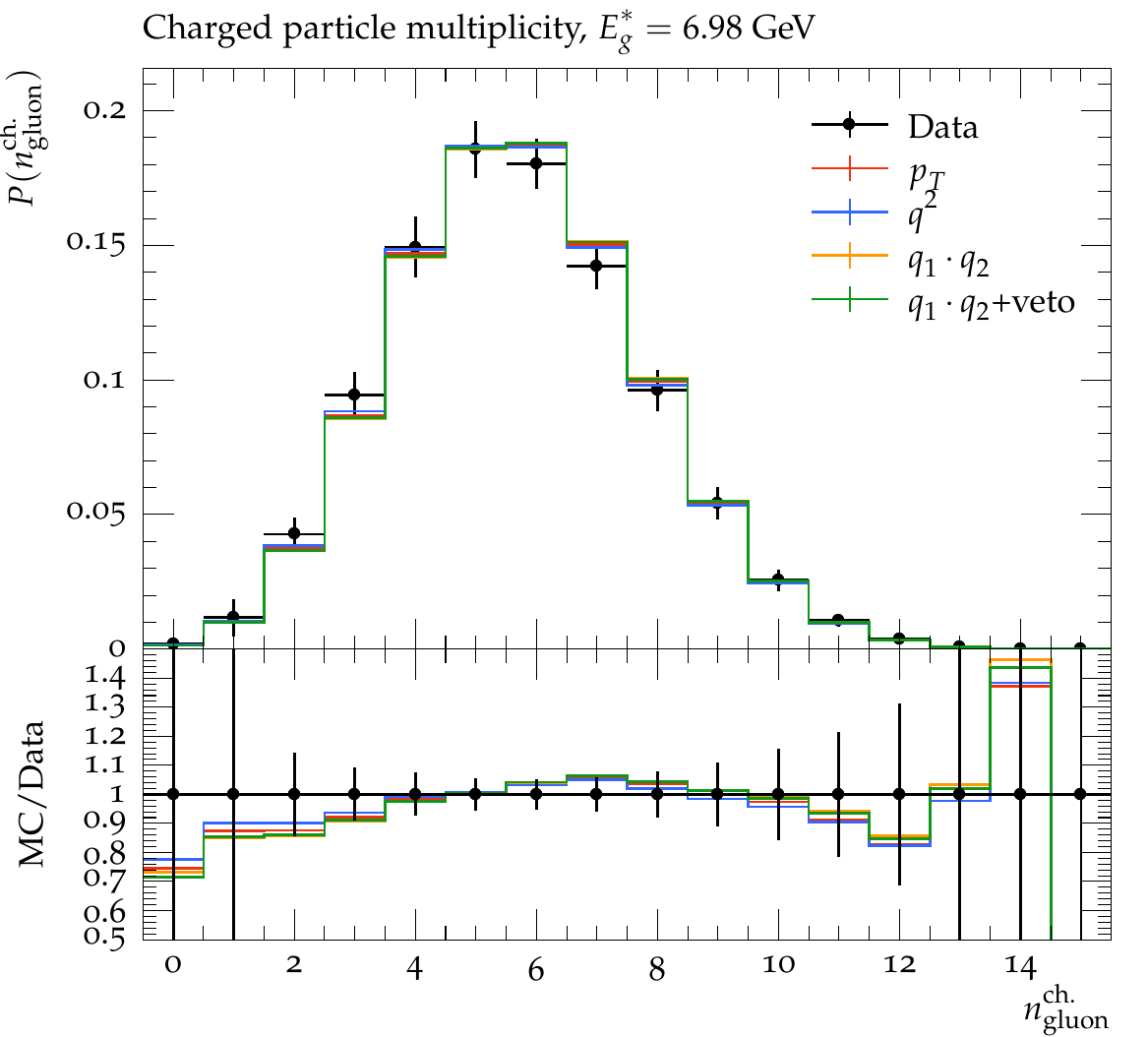}
\includegraphics[width=0.45\textwidth]{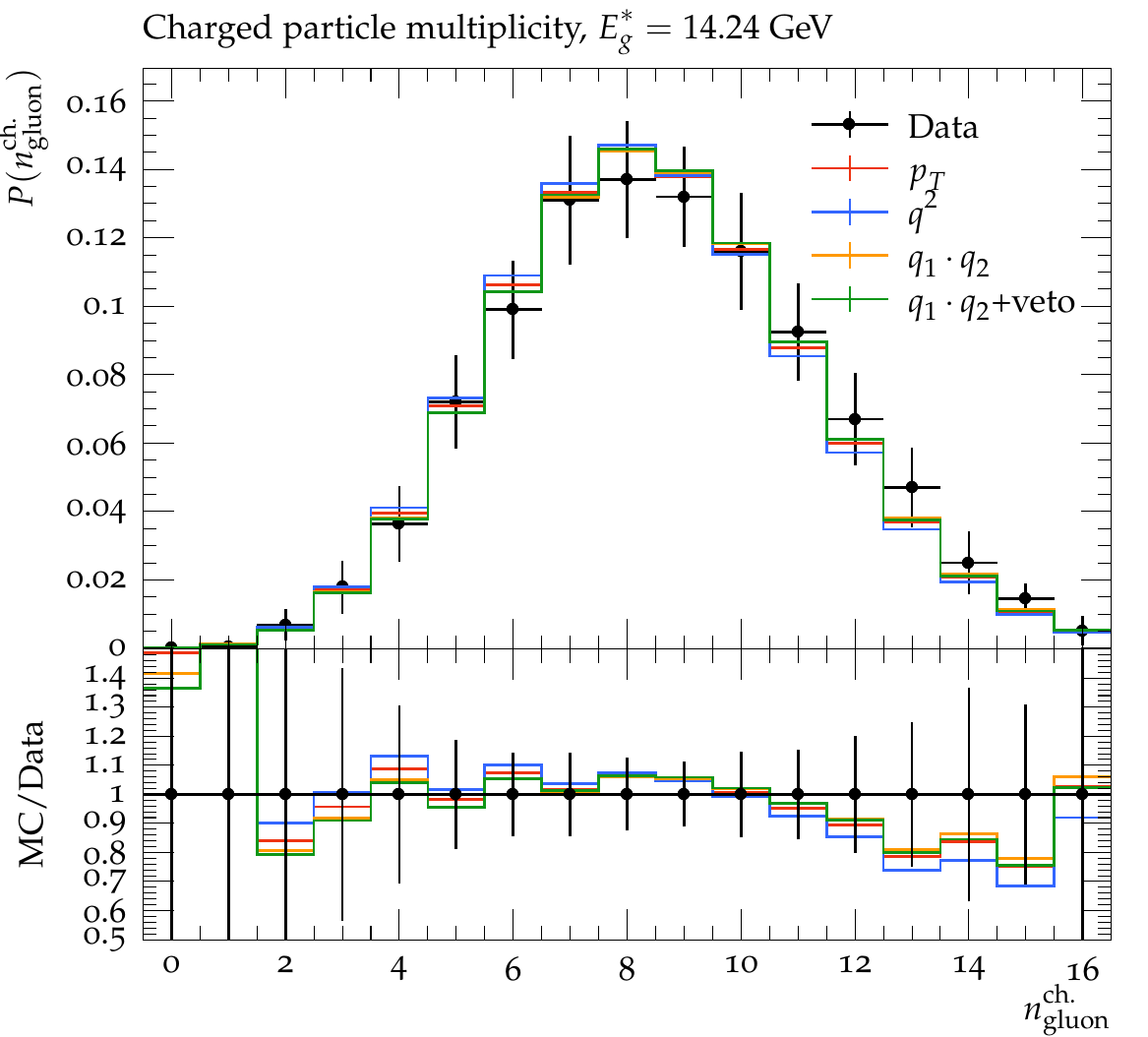}
\caption{Multiplicity distribution of charged particles in gluons jets for two different gluon energies compared with data from OPAL~\cite{Abbiendi:2003gh}.}
\label{fig:charged}
\end{figure}

In Fig.~\ref{fig:charged} we show the multiplicity distribution of charged particles in gluon jets for two different gluon energies. We see that the differences between all of the recoil schemes are much smaller than the experimental error and in general they all give a good agreement with the data.

\begin{figure}
\centering
\includegraphics[width=0.45\textwidth]{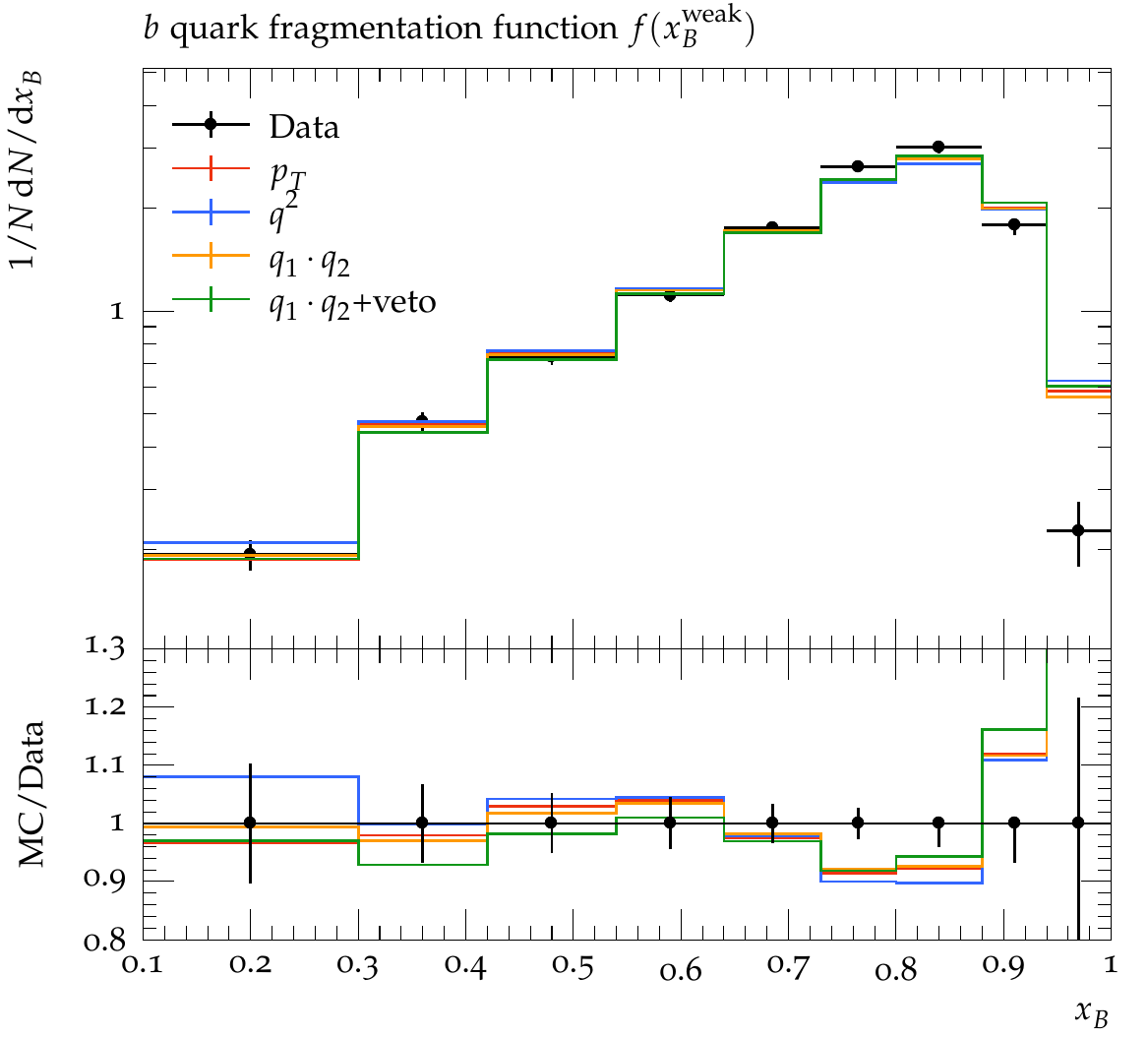}
\caption{Fragmentation function of weakly-decaying $B$-hadrons compared with data from DELPHI~\cite{DELPHI:2011aa}.}
\label{fig:bfrag}
\end{figure}

The schemes all fail to describe the peak region of the $b$
fragmentation function, with the different options making little
difference, see Fig.~\ref{fig:bfrag}. Nevertheless, the dot-product-plus-veto scheme gives the best overall description of $b$ data, as can
be seen from Tab.~\ref{tab:chi2}.

While all the data shown for $e^+e^-$ collisions was used as part of the tuning, this is true for all the tunes and therefore
the differences are due to the improvements in the parton shower.

\subsection{LHC results}

\begin{figure}
\includegraphics[width=0.45\textwidth]{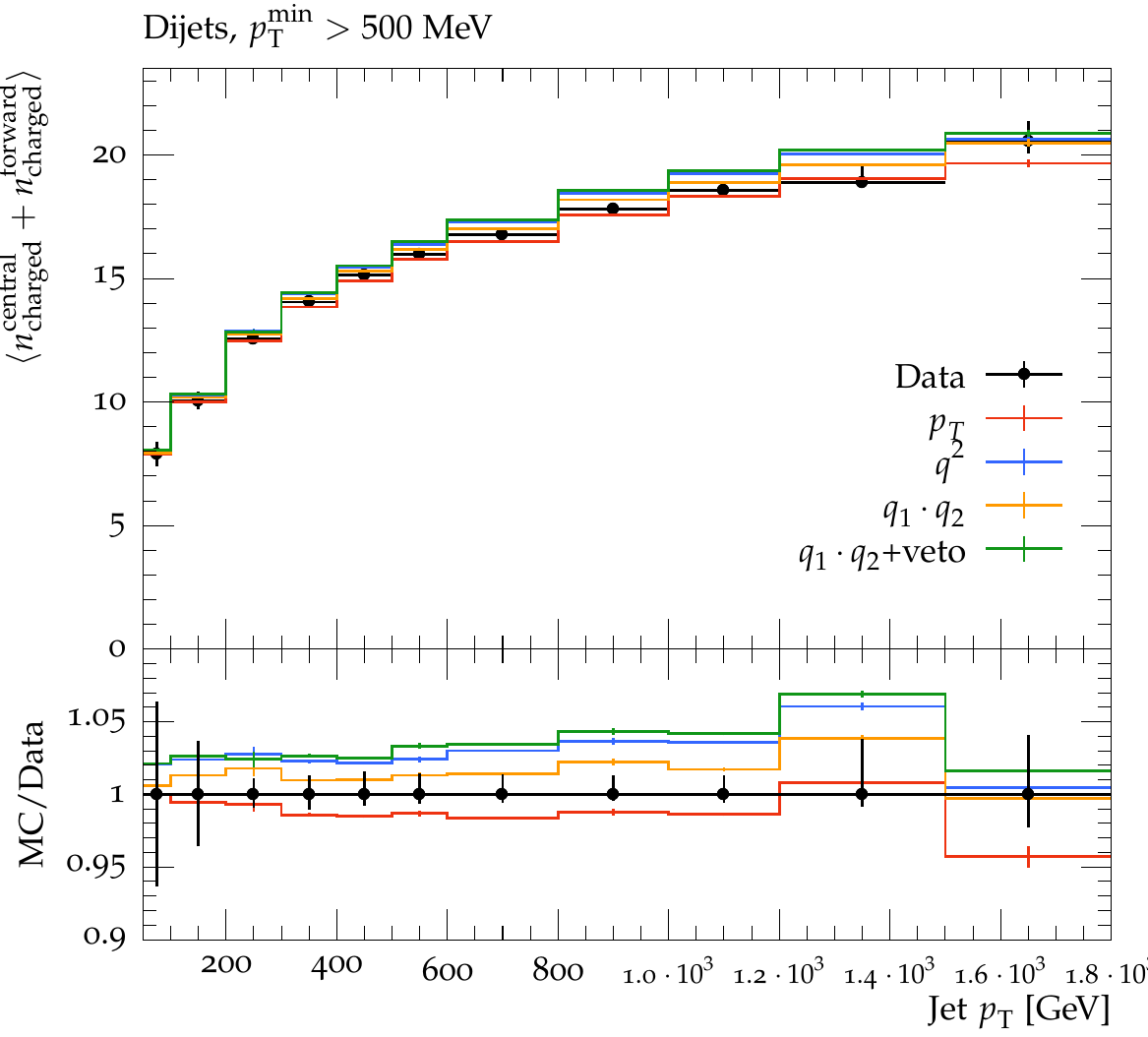}
\includegraphics[width=0.45\textwidth]{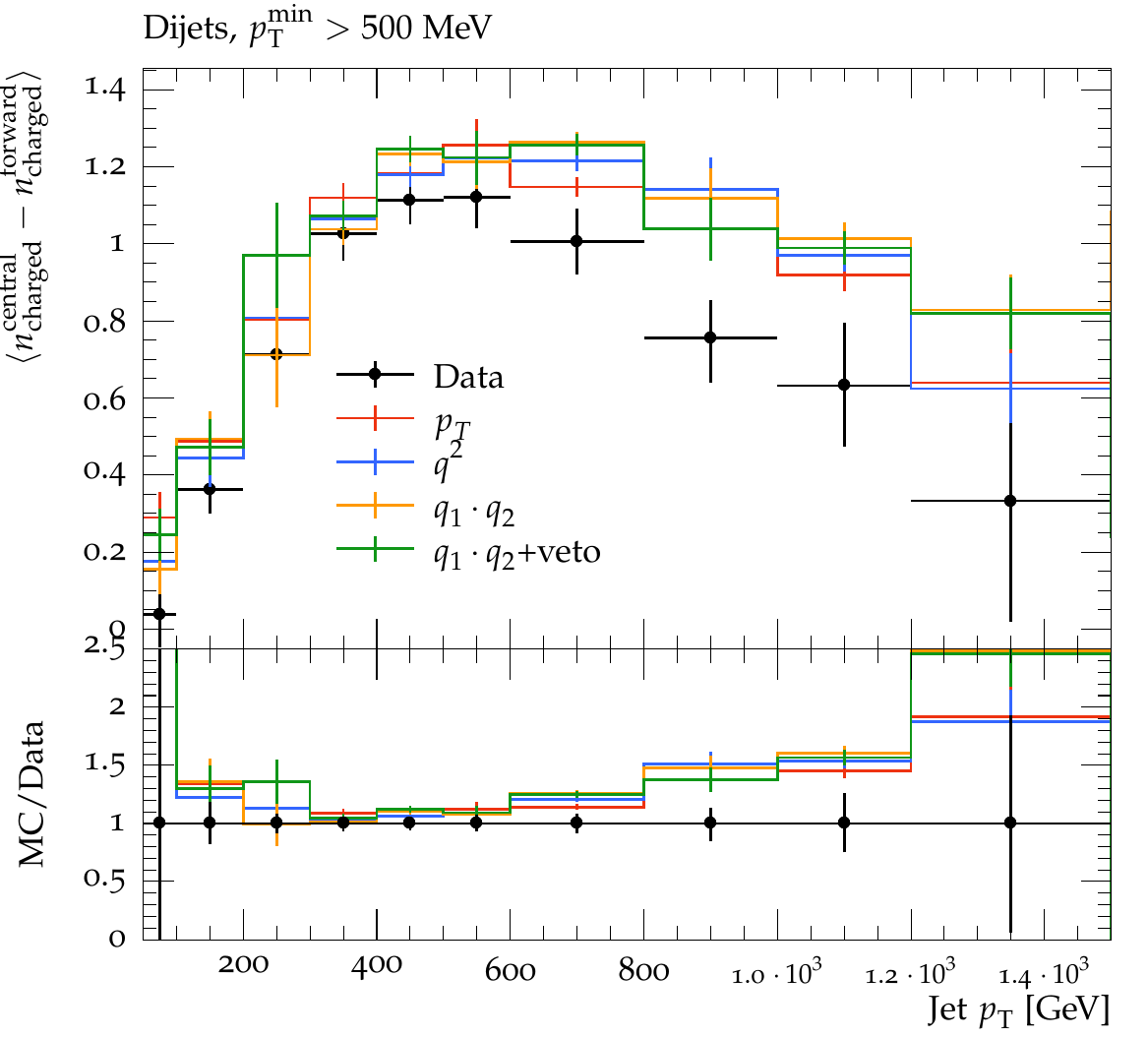}
\caption{The average number of charged particles in jets~(left) and the difference  between  the  average  number  of  particles  in central and forward jets~(right) as a function of the jet transverse momentum compared with data from the ATLAS experiment~\cite{Aad:2016oit}.}
\label{fig:atlas}
\end{figure}

Data from jets at the LHC seem to prefer the $p_T$ scheme as shown in Fig.~\ref{fig:atlas}. However, this behaviour is due to the absence of MEC in \Herwig{} for the events we are simulating. This implies that the dead zone remains unpopulated and the migration of events in this region partially solves the lack of hard emission generation. Nevertheless we do expect that matching with higher order computations will lead to the same behaviour that we find in LEP observables, \emph{i.e.}\ that the $p_T$ scheme yields too much hard radiation, while for the $q^2$ scheme, for which the kinematics of subsequent soft emissions are not guaranteed to be independent, we expect worse behaviour in the opposite region of the spectrum, and the dot-product-preserving scheme features intermediate properties. This data was not included in the tuning.
\vspace*{-2ex}% Improve page break. Might need removing if other changes are made

\section{Conclusions}
 
The pioneering work in Ref.~\cite{Dasgupta:2018nvj} investigated the
logarithmic accuracy of dipole showers by focusing on the pattern of
multiple emissions. Driven by this work, we have studied how different
choices of the recoil scheme in \Herwig{} can impact the logarithmic
accuracy of the distributions. 

We investigated the original choice of Ref.~\cite{Gieseke:2003rz},
where the transverse momentum of the emission is preserved during the
shower evolution, and the alternative proposal to preserve the
virtuality of the splitting, introduced in
Ref.~\cite{Reichelt:2017hts}. We observed that although the latter
prescription retains in general a good description of the experimental
data, it breaks the formal logarithmic accuracy of the parton shower, as
multiple soft emissions well separated in rapidity are not independent.
On the other hand, the older recoil scheme overpopulates the
non-logarithmically-enhanced region of the phase space, which should
not be filled by the parton shower, but instead by higher order
computations.

Due to the undesirable features of these recoil schemes, we proposed
an alternative interpretation of the angular-ordering variable that
well describes the process of multiple independent soft emission and
retains a good agreement with data while also considering the hard tail
of the distributions.  In order to enforce the correct behaviour in
the hard region of the spectrum, we implemented a veto that suppresses
large virtualities at the end of the parton shower. This veto applies
only to final state radiation and in the future we plan to propose an
extension which also includes initial state radiation. In the present
work we mainly focused on the case of a massless emitter. The study of
mass effects is crucial in assessing the accuracy of the parton shower
and will be considered in future works.

\acknowledgments

We thank our fellow \Herwig{} authors for useful
discussions, and Johannes Bellm and Bryan Webber for useful comments on the manuscript. We would also like to thank the referee for the many insights that helped us to improve the manuscript.
This work has received funding from the UK Science and Technology Facilities Council
(grant numbers ST/P000800/1, ST/P001246/1), the European Union’s Horizon 2020 research and innovation 
programme as part of the Marie Sk\l{}odowska-Curie Innovative Training Network MCnetITN3
(grant agreement no. 722104). GB thanks the UK Science and Technology
Facilities Council for the award of a studentship.

\appendix

\section{\boldmath$g\to q\bar{q}$ branching in the dot-product preserving scheme}
\label{app:gtoqqbar}

In the case of $g\to q\bar{q}$ branching the transverse momentum of the splitting, Eqn.\,\ref{eqn:qtilde_defn3}, becomes
\begin{eqnarray}
  p_T^2 &=& z^2(1-z)^2\tilde{q}^2  -(q_1^2-m^2)(1-z)^2 -(q_2^2-m^2)z^2 -m^2,
\end{eqnarray}
where $m$ is the quark mass.
So requiring
\begin{equation}
  \tilde{q}^2 > 2\max\left(\frac{q_1^2-m^2}{z^2}+\frac{m^2}{2z^2(1-z)^2},\frac{q_2^2-m^2}{(1-z)^2}+\frac{m^2}{2z^2(1-z)^2}\right),
\end{equation}
is sufficient, but not necessary, for there to be a physical solution in this case.
In this case the virtuality of the  branching is
 \begin{subequations}
   \begin{eqnarray}
 q_0^2 &=& q_1^2 + q_2^2 + z(1 - z)\tilde{q}^2 -2m^2 \leq \frac{\tilde{q}^2}2,
   \end{eqnarray}
 \end{subequations}
 which again will allow a solution but give a stricter limit.

\section{Impact of the recoil scheme on the logarithmic accuracy of the thrust distribution}
\label{sec:thrustLogs}
In this appendix we prove that the thrust is described only to LL accuracy in the $q^2$-preserving scheme, as this recoil scheme prescription introduces incorrect NLL terms at order $\alpha_s^2$.
To do so, we make use of the same methodology employed in Sec.~4 of Ref.~\cite{Dasgupta:2018nvj}, which relies on the CAESAR formalism~\cite{Banfi:2004yd}.
We introduce $\Sigma(L)$, which is the probability an event shape has a value smaller than $\exp(-L)$.
We have already seen in Sec.~\ref{sec:log_counting} that when we perform an expansion in the strong coupling, $\alpha_s$, at most 2 powers of $L$ appear for each power of $\alpha_s$, \emph{i.e.}
\begin{equation}
\Sigma(L) = \sum_{n=0}^\infty \sum_{m=0}^{2n} c_{m,n} \alpha_s^n L^m+\mathcal{O}(\alpha_s e^{-L})
\end{equation}
and therefore $\alpha_s^n L^{2n}$ are the LL contributions and $\alpha_s^n L^{2n-1}$ are the NLL ones.
For many event shapes, including the thrust, the expression for $\Sigma(L)$ can be rearranged to give
\begin{equation}
\Sigma(L) = \exp \left[ L g_1(\alpha_s L) + g_2 (\alpha_s L)+ \alpha_s g_3 (\alpha_s L) + \ldots \right] +\mathcal{O}(\alpha_s e^{-L}),
\end{equation} 
where the LL terms are contained in $g_1(\alpha_s L)$, while the NLL terms are in $ g_2 (\alpha_s L)$.

The Herwig single emission probability can be written as
\begin{equation}
 d P^{\texttt{Hw7}}_{\text{soft}} = \frac{2 \alpha_s C_F}{\pi} \frac{d \tilde{q}}{\tilde{q}} \frac{d \epsilon}{\epsilon} = \frac{2 \alpha_s C_F}{\pi} \frac{d p_T}{p_T} d \eta = \bar{\alpha}\frac{d p_T}{p_T} d \eta 
\end{equation}
where $\bar{\alpha}= \frac{2 \alpha_s C_F}{\pi}$, and $p_T = \epsilon \tilde{q}$ and $\eta = \log(\epsilon Q / p_T)$ are the Lund variables.
The impact of the incorrect shower mapping can be written as
\begin{align}
\delta \Sigma(L)  =& \bar{\alpha}^2\int_{-\infty}^{+\infty} d \eta_1 \int_{-\infty}^{-|\eta_1|} d \ell_1 \int_{-\infty}^{+\infty} d \eta_2 \, \int_{-\infty}^{-|\eta_2|} d \ell_2\, f(\eta_1,\eta_2) \int_0^{2\pi} \frac{d \phi_{12}}{2\pi} \\
& \times \left[ \Theta\big(e^{-L} - V_{\rm correct}(\eta_1, \ell_1,  \eta_2, \ell_2, \phi_{12})\big) - \Theta\big(e^{-L} - V_{\rm PS}(\eta_1, \ell_1,  \eta_2, \ell_2,\phi_{12})\big) \right],\nonumber
\end{align}
where we have replaced the $1/2!$ multiplicity factor with the ordering condition
\begin{equation}
f(\eta_1,\eta_2)
 = 
\begin{cases} 
 \Theta(\eta_2-\eta_1) & \mbox{if } \eta_1 \eta_2 <0  
 \\ \Theta(|\eta_2| - |\eta_1|)  & \mbox{if } \eta_1 \eta_2 >0 \end{cases}
\end{equation}
\emph{i.e.}\ either $\eta_1$ is in the left hemisphere and $\eta_2$ is in the right, or they are both in the same hemisphere and ordered with respect to each other.
$\ell_i = \log(p_{Ti}/Q)$ and $V$ is the shape observable expressed in terms of the Lund variables, the subscript ``correct'' means that $V$ is calculated using the correct double-soft kinematics where $p_{T1}\equiv\epsilon_1 \tilde{q}_1$ is the transverse momentum of the first emitted gluon, while ``PS'' denotes the result obtained using the kinematics of the \Herwig{} parton shower (in the double-soft limit).

In Sec.~\ref{sec:Lund} we have shown that the double-soft kinematics are correctly mapped if the transverse momenta or the dot products of the momenta of the emitted particles are preserved, so here
we only need to consider the case of the $q^2$-preserving scheme, which gives inaccurate kinematics when the two gluons are emitted from the same progenitor. We therefore only need to consider positive rapidities, provided we include a factor of 2
\begin{align}
\delta \Sigma(L)  =& 2 \bar{\alpha}^2\int_{0}^{+\infty} d \eta_1 \int_{-\infty}^{-\eta_1} d \ell_1 \int_{0}^{+\infty} d \eta_2 \, \int_{-\infty}^{-\eta_2} d \ell_2\, \Theta(\eta_2 - \eta_1) \int_0^{2\pi} \frac{d \phi_{12}}{2\pi}  \\
& \times \left[ \Theta\big(e^{-L} - V_{\rm correct}(\eta_1, \ell_1,  \eta_2, \ell_2,\phi_{12})\big) - \Theta\big(e^{-L} - V_{\rm PS}(\eta_1, \ell_1,  \eta_2, \ell_2,\phi_{12})\big) \right]. \nonumber
\end{align}
The correct expression for the thrust is
\begin{equation}
1-T = \frac{p_{T1} e^{-\eta_1} + p_{T2} e^{-\eta_2}}{Q} = \frac{p_{T1}^2}{\epsilon_1 Q^2} +  \frac{p_{T2}^2}{\epsilon_2 Q^2} = e^{\ell_1-\eta_1} + e^{\ell_2-\eta_2}.
\label{eq:Texact}
\end{equation}
In the case of the $q^2$-preserving scheme the contribution of the first gluon is modified: we label the new transverse momentum and rapidity as $\overline{p}_{T1}$ and $\overline{\eta}_{1}$ respectively, while we denote by $p_{T1}$ and $\eta_1$ the original values. Therefore from Eqn.~\eqref{eqn:qtilde2_pT} we can read that
\begin{equation}
p_{T1}^2 \to \overline{p}_{T1}^2 = \max \left( p_{T1}^2 - \frac{\epsilon_1}{\epsilon_2}p_{T2}^2, 0 \right).
\end{equation}
By observing that the recoil prescription does not change the light-cone momentum fraction of the first gluon, \emph{i.e.}
\begin{equation}
\epsilon_1 = \frac{\overline{p}_{T1}}{Q} e^{\overline{\eta}_{1}} = \frac{p_{T1}}{Q} e^{\eta_1},
\end{equation}
we can write
\begin{align}
1-T =&\;  \frac{\overline{p}_{T1}e^{-\overline{\eta}_{1}}}{Q}  + \frac{p_{T2} e^{-\eta_2}}{Q} = 
\frac{\overline{p}_{T1}^2 }{\epsilon_1 Q^2} + \frac{p_{T2}^2}{\epsilon_2 Q^2}  \nonumber \\
=&\; \max \left(\frac{p_{T1}^2}{\epsilon_1 Q^2} , \frac{p_{T2}^2}{\epsilon_2 Q^2} \right) =  \max(e^{\ell_1-\eta_1}, e^{\ell_2-\eta_2}).
\label{eq:Tq2}
\end{align}
By comparing Eqn.~\eqref{eq:Texact} and Eqn.~\eqref{eq:Tq2}, we notice that the two expressions coincide in the strongly ordered region, thus we expect the effect of the incorrect kinematic mapping to show only at NLL.
By performing the calculation we indeed find that
\begin{align}
\delta \Sigma(L)  =&\;  \bar{\alpha}^2\int_{0}^{+\infty} d \eta_1 \int_{-\infty}^{-\eta_1} d \ell_1 \int_{0}^{+\infty} d \eta_2 \, \int_{-\infty}^{-\eta_2} d \ell_2\, \int_0^{2\pi} \frac{d \phi_{12}}{2\pi} \nonumber \\
& \times \left[ \Theta\big(e^{-L} - e^{\ell_1-\eta_1} - e^{\ell_2-\eta_2}\big) - \Theta\big(e^{-L} - \max(e^{\ell_1-\eta_1}, e^{\ell_2-\eta_2} )\big) \right] \nonumber  \\
=&\;  2\bar{\alpha}^2\int_{0}^{\infty} d x_1 \int_{0}^{(L+x_1)/2} d \eta_1 \int_{x_1}^{\infty} d x_2 \int_{0}^{(L+x_2)/2} d \eta_2 \left[ \Theta(1 -e^{-x_1} - e^{-x_2}) -1 \right],
\end{align}
where in the first line we have removed the theta function coming from the angular-ordering condition, $\Theta(\eta_2 - \eta_1)$, and included a factor of $1/2$ as the integrand is symmetric in the exchange $1\leftrightarrow 2$.
In the second line we have defined $x_i=\eta_i-\ell_i-L$ and reinserted an ordering $x_2>x_1$. Now, the only dependence on $L$ is in the limits on the $\eta$ integrals, which are trivial, and we can read off the leading power in $L$,
\begin{align}
=&\;  -\frac{\bar{\alpha}^2}{2}L^2\int_{0}^{\infty} d x_1 \int_{x_1}^{\infty} d x_2 \; \Theta(e^{-x_1} + e^{-x_2} - 1) + \mathcal{O}(\bar{\alpha}^2L) \nonumber \\
=&\;  -\frac{\bar{\alpha}^2}{2}L^2\int_{0}^{\log2} d x_1 \left[-\log(e^{x_1}-1)\right] + \mathcal{O}(\bar{\alpha}^2L) \nonumber \\
=&\; -\frac{ \bar{\alpha}^2}{2} \frac{\pi^2}{12} L^2 + \mathcal{O}(\bar{\alpha}^2L) \nonumber \\
=&\; - \frac{C^2_F}{6} \alpha_s^2 L^2 + \mathcal{O}(\alpha_s^2L),
\end{align}
This proves that this choice of the kinematic mapping introduces a NLL discrepancy at order $\alpha_s^2$ (while in the case of dipole showers, the first NLL discrepancy appears at order $\alpha_s^3$~\cite{Dasgupta:2018nvj}). 

\clearpage
\bibliography{herwig}
\end{document}